\documentclass[conference]{IEEEtran}
\IEEEoverridecommandlockouts
\usepackage{cite}
\usepackage{amsmath,amssymb,amsfonts}
\usepackage{graphicx,color,overpic,psfrag}
\usepackage{xcolor}
\usepackage{algorithmic}
\usepackage{graphicx}
\usepackage{textcomp}
\usepackage{fancyhdr}
\usepackage{xcolor}
\usepackage{graphicx}
\usepackage{bm}
\usepackage{subfigure}
\usepackage{multirow}
\usepackage{booktabs}
\usepackage{caption}
\usepackage{booktabs}
\usepackage{amsmath}
\usepackage[numbers,sort&compress]{natbib}
\usepackage{algorithm}
\usepackage{algorithmic}
\usepackage{amsfonts}
\usepackage{hyperref}
\usepackage{xcolor}
\usepackage[table]{xcolor}
\usepackage{amsmath}
\usepackage{tabularx}
\usepackage{threeparttable}
\hypersetup{
  colorlinks=true,
  linkcolor=blue,  
  citecolor=blue,  
  urlcolor=blue,   
  linkbordercolor={0 0 1},
  pdfborderstyle={/S/U/W 1} 
}

\def\BibTeX{{\rm B\kern-.05em{\sc i\kern-.025em b}\kern-.08em
    T\kern-.1667em\lower.7ex\hbox{E}\kern-.125emX}}

\newcommand{\bl}[1]{\color{black}#1}

\begin{document}

\title{Vision-Based Efficient Joint Trajectory and Channel Tracking in Near-Field XL-MIMO Systems

\author{Mengyuan Li, \textit{Graduate Student Member, IEEE}, Yu Han, \textit{Member, IEEE}, Hao Xu, \textit{Senior Member, IEEE},\\Yongxu Zhu, \textit{Senior Member, IEEE},  Shi Jin, \textit{Fellow, IEEE}, and Chao-Kai Wen, \textit{Fellow, IEEE}\\
}
\thanks{M. Li, Y. Han, H. Xu, S. Jin, and Y. Zhu are with the School of Information Science and Engineering, Southeast University, Nanjing 210096, China (email: mengyuan$\_$li@seu.edu.cn; hanyu@seu.edu.cn; hao.xu@seu.edu.cn; yongxu.zhu@seu.edu.cn; jinshi@seu.edu.cn). 

C.-K. Wen is with the Institute of Communications Engineering, National Sun Yat-sen University, Kaohsiung 804, Taiwan (e-mail: chaokai.wen@mail.nsysu.edu.tw).
}
}
\maketitle

\begin{abstract}
Accurate joint tracking of mobile users, surrounding scatterers, and dynamic channels is a critical task for sixth-generation (6G) wireless systems, essential for both ensuring high-quality communications and empowering advanced selsing applications such as autonomous driving and immersive extended reality. While extremely large-scale multiple-input multiple-output (XL-MIMO) inherently offers strong support for this task through its high spatial resolution and spectral efficiency, its massive scale of antenna arrays, coupled with near-field propagation characteristics, makes joint trajectory and channel tracking time-consuming and hardware-intensive. {\bl To address these challenges, we rethink the problem from a vision-based signal perspective. Specifically, we design a subarray-based partially connected hybrid beamforming (PC-HBF) architecture with a tailored time-multiplexed (TM) mechanism. This effectively compensates for the aperture loss caused by limited radio frequency (RF) chains, generating high-fidelity Cartesian-domain signal images that inherently capture near-field spatial features. Based on this visual representation, we propose an improved CenterNet to perform accurate one-shot path localization, circumventing the path-iterative search required by conventional compressed-sensing-based methods. Building upon this to further improve the accuracy and exploit temporal correlation, a local small-scale orthogonal matching pursuit (OMP) refiner and a lightweight cascaded OMP tracker are developed. Finally, a Hungarian-based trajectory association module is incorporated to maintain track continuity and provide trajectory-level information for environment monitoring.} Simulation results show that the proposed framework consistently outperforms representative baselines in position and channel tracking accuracy, especially under low-SNR and limited-hardware conditions.
\end{abstract}


\section{Introduction} 

\IEEEPARstart{I}{n} future sixth-generation (6G) wireless systems, emerging applications such as autonomous driving, smart transportation, and industrial automation require wireless networks to provide not only high data rates and low latency, but also reliable sensing ability of the user and dynamic environments~\cite{Wang_SCIS_6G_Vision}. As a key enabling technology for 6G, extremely large-scale multiple-input multiple-output (XL-MIMO) provides the foundation for such services by offering unprecedented spatial resolution, strong near-field beam focusing, and improved spectral and energy efficiency~\cite{XL_MIMO_Tutorial,Lu_COMST_XLMIMO_Tutorial}. Benefiting from the spherical wavefront of near-field channels, XL-MIMO systems inherently embed both angle and distance information of users and scatterers into the channel, which enables high-resolution propagation path localization.

Existing studies have demonstrated the potential of XL-MIMO for joint localization and channel estimation by exploiting near-field angle-distance coupling codebooks through compressed-sensing-based path-iterative search-and-refinement methods~\cite{tian_vr,NF_Localization_Joint_CE,nnomp1,Xi_Gridless_WCL2024,SIGW,ScattererLocalization_Spherical,ScattererLocalization_SAGE}. To further reduce the computational complexity introduced by the large codebook, some studies have also explored structured channel representations, e.g., transforming antenna-domain channels into angle-distance-domain signal images so that all propagation paths can be extracted from these images efficiently through one-shot detection network inference~\cite{keypointCE}.

{\bl While accurate estimation at the initial time slot is crucial, efficient subsequent tracking also relies heavily on exploiting temporal correlation and the structured evolution of propagation paths. Existing tracking approaches can be broadly classified into two categories. The first is \emph{Bayesian recursive tracking}, where time-varying propagation parameters are modeled as latent states and sequentially updated from pilot observations using recursive Bayesian inference engines, such as extended or unscented Kalman filters (EKF/UKF)~\cite{Zhao_TWC_STBEM_2018, Zhao_TCOM_AngleTracking_2017}. The second is \emph{path-parameter-based tracking}, where methods such as Newtonized orthogonal matching pursuit (NOMP) explicitly exploit geometric path relationships to extrapolate propagation states over time~\cite{extendedNOMPbasedTracking}. 
As systems scale to XL-MIMO, these tracking methods have been extended to more challenging propagation regimes. On the one hand, Bayesian recursive tracking has been generalized to high-dimensional near-field scenarios using sparse or hierarchical Bayesian inference schemes~\cite{Guerra_TSP_NFTracking_2021,Chen_Arxiv_HybridTracking_2025,Yuan_WCNC_XLRIS_2024,Tuo_XLIRS_SparseBayesian_2025}. On the other hand, path-parameter-based methods have also been adapted to exploit dynamic characteristics of geometric paths and spatial non-stationarity in near-field channels~\cite{Xu_DynamicSparsity_2024,Zhang_TWC_NFTimeVarying_2026,Joint_VR_CE_XLMIMO,Integrated_LocSense_Comm}. 

Despite these advances, near-field XL-MIMO tracking remains fundamentally more challenging than its far-field counterpart. Due to the spherical-wave propagation characteristic, angle and distance become strongly coupled in the steering vector, leading to highly nonlinear and often ill-conditioned parameter estimation problems. For Bayesian recursive trackers, this coupling complicates the state-observation mapping, so linearization-, sampling-, or posterior-inference-based updates may suffer from model mismatch, error propagation, or even instability. For path-parameter-based methods, the enlarged angle-distance search space makes path-iterative search-and-refinement increasingly expensive. As the observation area expands or the multipath components increase, this poor scalability leads to severe latency issues, precluding their use in practical systems.
}

Beyond these algorithmic challenges, the practical deployment of XL-MIMO also requires careful consideration of hardware constraints.
Owing to the extremely large array size, fully digital beamforming requires one RF chain per antenna and thus incurs prohibitive hardware cost and power consumption. 
Hybrid beamforming (HBF) becomes a practical choice for XL-MIMO implementations~\cite{Zhu_Subarray_WCL2023,Wang_XLMIMO_Mag_2024}. 
Although recent studies have investigated HBF designs for XL-MIMO systems~\cite{deepunfolding_hybridBF,hybrid_ris1}, HBF for the near-field sensing and tracking is far from a transparent front-end compression module. Consequently, how to tailor the HBF design to facilitate efficient tracking from the received signals remains underexplored.

\begin{table}[t]
\centering
\caption{{\bl Comparison of the proposed framework with representative tracking methods across different categories.}}
\label{tab:novelty_comparison}
\renewcommand{\arraystretch}{1.2}
\setlength{\tabcolsep}{4pt}
\footnotesize
\resizebox{0.48\textwidth}{!}{%
\begin{tabular}{lcc}
    \toprule
    \textbf{Tracking Paradigm} & \textbf{Iterative Inference} & \textbf{HBF-Compatible} \\
    \midrule
    Bayesian Recursive Tracking \cite{Zhao_TWC_STBEM_2018, Zhao_TCOM_AngleTracking_2017, Guerra_TSP_NFTracking_2021,Chen_Arxiv_HybridTracking_2025,Yuan_WCNC_XLRIS_2024,Tuo_XLIRS_SparseBayesian_2025} 
    & \checkmark & Partial \\
    
    Path-Parameter-Based Tracking \cite{extendedNOMPbasedTracking, Xu_DynamicSparsity_2024,Zhang_TWC_NFTimeVarying_2026,Joint_VR_CE_XLMIMO,Integrated_LocSense_Comm} 
    & \checkmark & $\times$ \\
    
    \rowcolor{gray!15}
    \textbf{Proposed Vision-Based} 
    & \textbf{$\times$} & \textbf{\checkmark} \\
    \bottomrule
\end{tabular}%
}
\end{table}

{\bl
In this paper, we propose a vision-based framework for efficient joint trajectory and channel tracking in time-varying near-field XL-MIMO systems. As summarized in Table~\ref{tab:novelty_comparison}, our approach establishes a streamlined “signal-to-imag” paradigm. By synergizing hardware-efficient signal-image acquisition, fast learning-driven one-shot path detection, and robust model-based temporal tracking, the proposed framework bypasses massive computational overhead, enabling highly accurate and low-complexity tracking for dynamic channels.

\textbf{Contributions:} The main contributions of this paper are summarized as follows:

\begin{itemize}
\item \textit{Tailored Hybrid Beamforming Design for Received Signal Image Generation:} To address the aperture loss under strict RF chain constraints, we introduce a time-multiplexed (TM) signal combining strategy within a partially connected HBF architecture. This HBF scheme can successfully generate high-quality signal images even when the number of RF chains is severely limited, establishing a hardware-efficient foundation for accurate tracking.

\item \textit{Learning-Based Fast Path Detection and Refinement:} We design an improved CenterNet detector to extract all path locations rapidly from the received signal images through a single inference, thereby avoiding path-iterative search-and-refinement in conventional compressed sensing methods. Building upon this, a lightweight local OMP refiner and a least-squares (LS) update are integrated to further improve the accuracy of path and channel estimations.

\item \textit{Geometry-Aware Tracking and Environment Monitoring:} To efficiently track the trajectory and channel, we develop an efficient cascaded OMP tracker that exploits geometric temporal correlations, instead of estimating for all time slots. This is augmented by a residual-based validity check to seamlessly accommodate dynamic path appearances and disappearances. Finally, a Hungarian-based association module links these per-slot geometric estimates into continuous trajectories, supporting trajectory-level dynamic environment interpretation.

\end{itemize}
}


{\bf Notations.} Bold uppercase (lowercase) letters denote matrices (vectors). $(\cdot)^{\top}$, $(\cdot)^H$, and $(\cdot)^{\dagger}$ represent transpose, conjugate transpose, and pseudo-inverse, respectively. $\odot$, $\mathbb{E}\{\cdot\}$, $|\cdot|$, and $\|\cdot\|$ denote Hadamard product, expectation, absolute value, and Euclidean norm. $\lceil \cdot \rceil$ denotes the ceiling function. For vector $\mathbf{x}$, $|\mathbf{x}|_{\min}$ ($|\mathbf{x}|_{\max}$) is the minimum (maximum) entry magnitude. $\mathbf{I}_N$ denotes the $N \times N$ identity matrix. $\mathrm{blkdiag}(\cdot)$ constructs a block-diagonal matrix. $\mathbf{A}[i_1\!:\!i_2, j_1\!:\!j_2]$ denotes the submatrix spanning rows $i_1+1$ to $i_2$ and columns $j_1+1$ to $j_2$. $\mathcal{CN}$ is the complex Gaussian distribution. $\mathbb{Z}^{+}$ denotes the set of positive integers.

\section{System Model}
\label{sec:System Model}

In this section, we introduce the considered time-varying XL-MIMO system and the problem formulation.

\subsection{System Model}

We consider an uplink XL-MIMO terrestrial communication system in urban traffic scenarios, where the base station (BS) is equipped with an $N$-element uniform linear array (ULA) with antenna spacing $\lambda/2$, operating at central carrier frequency $f_{\mathrm{c}}$. The wavelength is denoted as $\lambda=c/f_{\mathrm{c}}$, where $c=3\times 10^8$~m/s. {\bl In this considered system, both the user equipment (UE) and the surrounding scatterers are located in the established ground x-z plane, specifically confined within an observed region bounded by $x \in [x_{\min}, x_{\max}]$ and $z \in [z_{\min}, z_{\max}]$.}

\begin{figure}[t]
\centering
\includegraphics[width=0.75\linewidth]{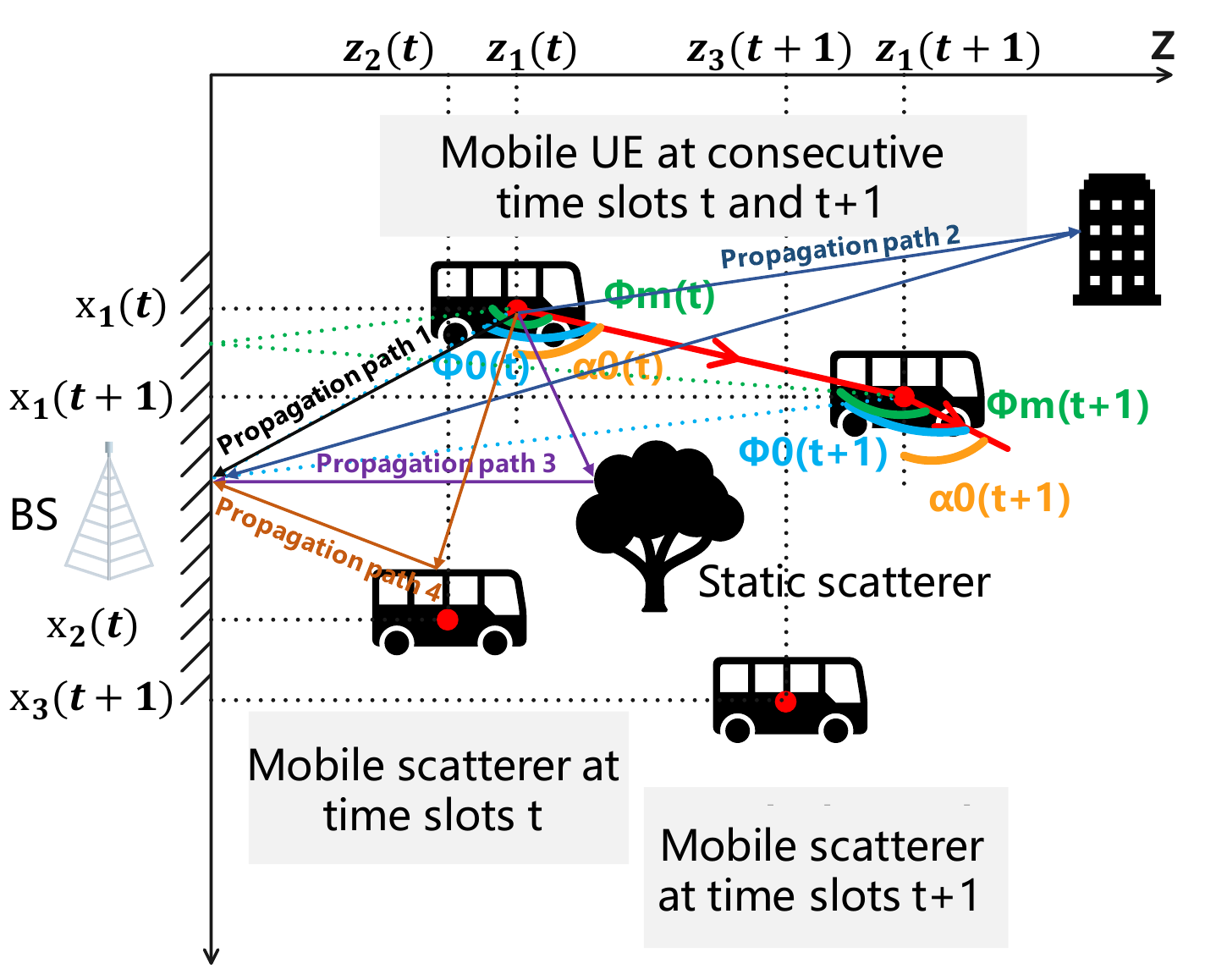}
\caption{{\bl Illustration of the considered time-varying XL-MIMO system, including one mobile user and surrounding static and mobile scatterers distributed in the observed near-field region.}}
\label{fig:system_model}
\end{figure}

As illustrated in Fig.~\ref{fig:system_model}, the propagation environment consists of one UE and a set of surrounding scatterers. We use the index $l$ to enumerate the propagation paths. {\bl Specifically, the UE corresponds to $l=1$ with position $\big(z_1(t),x_1(t)\big)$, representing the line-of-sight (LoS) path.} The $L_1$ static scatterers are indexed as
\begin{equation}
\mathcal{L}_{\mathrm{sta}} \triangleq \{2,3,\ldots,L_1+1\},
\end{equation}
which are assumed to remain unchanged during the observation interval. Moreover, there are $L_2(t)$ mobile scatterers at time slot $t$, indexed as
\begin{equation}
\mathcal{L}_{\mathrm{mob}}(t) \triangleq \{L_1+2,\ldots,L_1+1+L_2(t)\},
\end{equation}
with time-varying positions $\big(z_l(t),x_l(t)\big)$. Accordingly, the total number of propagation paths at time slot $t$ is
\begin{equation}
L(t)=1+L_1+L_2(t).
\end{equation}



{\bl The red arrows in Fig.~\ref{fig:system_model} indicate the motion directions of the mobile UE and mobile scatterers. Specifically, $\alpha_l(t)$ denotes the motion-direction angle of the $l$-thuser/scatterer relative to the x-axis, whereas $\phi_m(t)$ denotes the corresponding geometric angle associated with the $m$-th antenna element.}

For the UE ($l=1$) and each mobile scatterer ($l\in\mathcal{L}_{\mathrm{mob}}(t)$), we adopt the following Markovian motion model:
\begin{equation}
\begin{bmatrix}
z_l(t) \\ x_l(t)
\end{bmatrix}
=
\begin{bmatrix}
z_l(t-1) \\ x_l(t-1)
\end{bmatrix}
+
v_l(t)T_s
\begin{bmatrix}
\sin\!\big(\alpha_l(t)\big) \\
\cos\!\big(\alpha_l(t)\big)
\end{bmatrix},
\label{eq:motion_model}
\end{equation}
where $T_s$ is the slot duration, $v_l(t)$ is the instantaneous speed satisfying $v_{\min}\le v_l(t)\le v_{\max}$, and $\alpha_l(t)$ denotes the moving direction.
Moreover, the mobile scatterer follows a birth-death process. At each time slot, new mobile scatterers appear with probability $P_{n}$, and each existing mobile scatterer disappears independently with probability $P_{d}$. As a result, $L_2(t)$ and the associated index set $\mathcal{L}_{\mathrm{mob}}(t)$ vary over time.

\subsection{Channel Model}

The resulting time-varying near-field uplink channel at time slot $t$ can be modeled as
\begin{equation}
\mathbf{h}(t)
=
\sum_{l=1}^{L(t)} g_l(t)\,\mathbf{a}_l(t)\odot \mathbf{d}_l(t), \quad \mathbf{h}(t)\in\mathbb{C}^{N\times 1},
\label{eq:channel_model}
\end{equation}
where $g_l(t)$ denotes the complex gain of the $l$-th path. The steering vector $\mathbf{a}_l(t)\in \mathbb{C}^{N\times 1}$ can be expressed as
\begin{equation}
\mathbf{a}_l(t)
=
\left[
\tfrac{1}{r_{\frac{1-N}{2}}(t)} e^{j k_c r_{\frac{1-N}{2}}(t)},\,
\dots,\,
\tfrac{1}{r_{\frac{N-1}{2}}(t)} e^{j k_c r_{\frac{N-1}{2}}(t)}
\right]^{H},
\label{eq:steering_vector}
\end{equation}
where the wave number $k_c=2\pi/\lambda$, $r_n(t)$ denotes the distance from the $l$-th propagation path to the $n$-th antenna element ($n={(1-N)/2},\ldots,{(N-1)/2}$), and can be represented as
\begin{equation}
r_n(t)
=
\sqrt{z_l^2(t) + \big(x_l(t) - n \lambda / 2\big)^2}.
\end{equation}
 {\bl $1/r_n(t)$ accounts for the free-space path loss attenuation inherent to spherical wavefront propagation in the near-field region.}
The Doppler phase-shift vector $\mathbf{d}_l(t)\in \mathbb{C}^{N\times1}$ is
\begin{equation}
\begin{aligned}
\mathbf{d}_l(t) = \Big[ & e^{j k_c v_l(t) T_s \cos \!\big(\phi_{l,\frac{1-N}{2}}(t)\big)}, \\
& \dots,\, e^{j k_c v_l(t) T_s \cos \!\big(\phi_{l,\frac{N-1}{2}}(t)\big)} \Big]^{H}
\end{aligned}
\end{equation}
where $\phi_{l,n}(t)$ denotes the angle between the motion direction of the $l$-th path and the direction toward the $n$-th antenna.

{\bl
\textit{Remark 1 (Rationale of the Adopted Channel Model):}
The channel model in~\eqref{eq:channel_model} is adopted as a narrowband spatial abstraction to focus on the key near-field spatial characteristics, including spherical-wave propagation, angle-distance coupling, and time-varying geometric evolution. For a practical OFDM system, the frequency-domain channel on the $k$-th subcarrier can be written as
\begin{equation}
\mathbf h_k(t)
=
\sum_{l=1}^{L(t)}
g_l(t)e^{-j2\pi f_k\tau_l(t)}
\mathbf a_k(z_l(t),x_l(t))
\odot
\mathbf d_{l,k}(t),
\end{equation}
where $f_k=f_c+k\Delta f$ and $\tau_l(t)$ is the delay of the $l$-th path. For a fixed pilot subcarrier or a sufficiently narrow subband, the delay term can be absorbed into the effective gain $\tilde g_{l,k}(t)=g_l(t)e^{-j2\pi f_k\tau_l(t)}$, so the model reduces to the narrowband spatial form in~\eqref{eq:channel_model}. }

{\bl
\subsection{Problem Formulation}

To enable practical deployment, we consider a HBF architecture, where the BS is equipped with only $N_{\mathrm{RF}}$ RF chains and $N_{\mathrm{RF}}<N$. Let $\mathbf{W}\in\mathbb{C}^{N_{\mathrm{RF}}\times N}$ denote the analog combining matrix. Assuming the pilot symbol is 1, the compressed received signal $\mathbf{y}(t)\in\mathbb{C}^{N_{\mathrm{RF}}\times 1}$ is given by
\begin{equation}
    \mathbf{y}(t)=\sqrt{P_r}\mathbf{W}\mathbf{h}(t)+\tilde{\mathbf{n}}(t),
    \label{eq:general_signal_model}
\end{equation}
where $P_r$ denotes the pilot power and $\tilde{\mathbf{n}}(t)$ is the effective post-combining noise.

Given $\{\mathbf{y}(t)\}_{t=1}^{T}$, the considered task is to estimate the dominant path locations and complex gains, i.e.,
$\{(\hat z_l(t),\hat x_l(t)),\hat g_l(t)\}_{l=1}^{\hat L(t)}$,
and reconstruct the channel as
\begin{equation}
\hat{\mathbf h}(t)=\sum_{l=1}^{\hat L(t)} \hat g_l(t)\mathbf a(\hat z_l(t),\hat x_l(t)).
\end{equation}
Accordingly, the overall task can be viewed as the following dynamic path-parameter estimation problem:
\begin{small}
\begin{equation}
\min_{\substack{\{L(t)\}_{t=1}^{T},\\ \{z_l(t),x_l(t),g_l(t)\}}}
\sum_{t=1}^{T}
\left\|
\mathbf y(t)-\sqrt{P_r}\mathbf W
\sum_{l=1}^{L(t)} g_l(t)\mathbf a\big(z_l(t),x_l(t)\big)
\right\|_2^2.
\label{eq:overall_objective}
\end{equation}
\end{small}
However, \eqref{eq:overall_objective} is highly nonconvex due to the coupled geometric parameters, the time-varying path set, and the temporal evolution across slots. Therefore, instead of solving it directly, we adopt a decomposed coarse-to-fine framework with two stages: initial path estimation at $t=1$ and subsequent geometry-aware path tracking for $t=2,\ldots,T$.

Specifically, the first stage recovers the dominant near-field paths from $\mathbf y(1)$ to initialize the UE/scatterer positions, gains, and channel. The second stage exploits the temporal correlation of the previously estimated paths to update $\{(\hat z_l(t),\hat x_l(t)),\hat g_l(t)\}_{l=1}^{\hat L(t)}$ sequentially over time, thereby enabling continuous channel reconstruction as well as dynamic event identification, such as path appearance, disappearance, and abrupt motion changes. The practical structural constraints are imposed by the hybrid beamforming architecture through $\mathbf{W}$ and the limited RF-chain budget.
}

\begin{figure*}[t]
    \centering
        \includegraphics[width=0.75\linewidth]{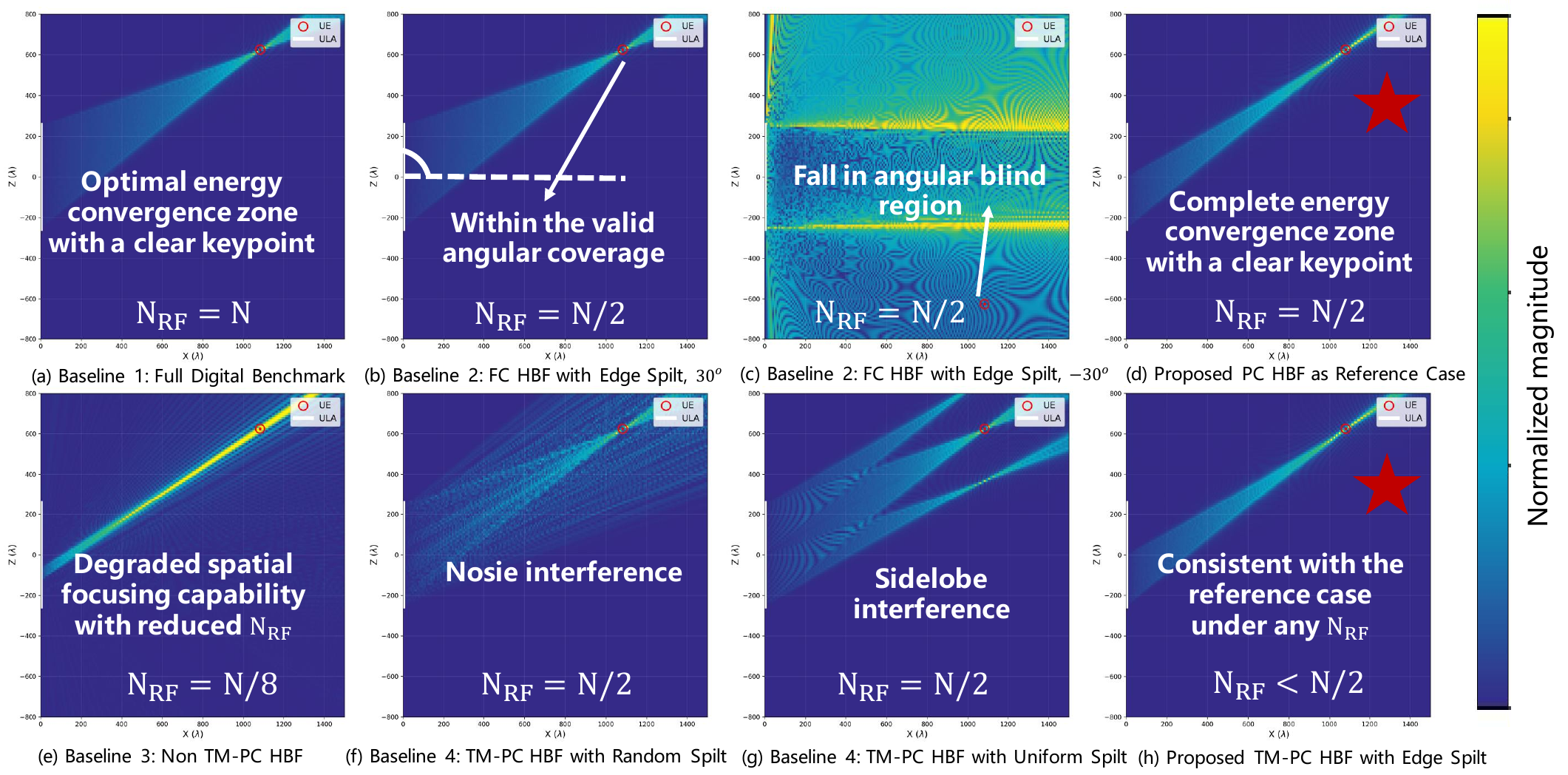} 
\caption{{\bl Comparison of generated signal images under various beamforming schemes. 
}}

    \label{fig:channel_images_rgb}
\end{figure*}

\section{HBF Design for High-Quality Received Signal Image Generation}
\label{sec:Hybrid Beamforming Channel Image Generation}

In this section, we develop a tailored HBF scheme to generate the received signal images that can preserve distinct propagation path features with limited RF chains.
\subsection{Signal Image Formation and HBF Design Motivation}
\label{subsec:image_formation_motivation}

Ideally, under the full-digital beamforming architecture with $N_{\mathrm{RF}} = N$ and $\mathbf{W} = \mathbf{I}_N$ (\textit{Baseline 1}), the received signal $\mathbf{y}^{\mathrm{D}}(t) \in \mathbb{C}^{N \times 1}$ can be expressed by
\begin{equation}
\mathbf{y}^{\mathrm{D}}(t)
=
\sqrt{P_r}\,\mathbf{h}(t) + \mathbf{n}(t).
\end{equation}
{\bl A Cartesian-domain channel representation can be obtained by projecting $\mathbf{y}^{\mathrm{D}}(t)$ onto the corresponding global Cartesian codebook $\mathbf{U}_{\mathrm{T}}^{\mathrm{D}}$ as
\begin{equation}
\mathbf{y}_{\mathrm{C}}^{\mathrm{D}}(t)
=
\mathbf{U}_{\mathrm{T}}^{\mathrm{D}} \mathbf{y}^{\mathrm{D}}(t)
\in \mathbb{C}^{G_Z^{\text{gb}} G_X^{\text{gb}} \times 1},
\end{equation}
where $\mathbf{U}_{\mathrm{T}}^{\mathrm{D}}$ is constructed over the uniformly sampled Cartesian grid:
\begin{equation}
\mathcal{G}_{\text{gb}} = \big\{ (z_{\min} \!+\! (i\!-\!1)\Delta z,\, x_{\min} \!+\! (j\!-\!1)\Delta x) \big\},
\end{equation}
for $i=1,\ldots,G_Z^{\text{gb}}$ and $j=1,\ldots,G_X^{\text{gb}}$, with cardinality $|\mathcal{G}_{\text{gb}}|=G_Z^{\text{gb}}G_X^{\text{gb}}$, where $\Delta z=(z_{\max}-z_{\min})/(G_Z^{\text{gb}}-1)$ and $\Delta x=(x_{\max}-x_{\min})/(G_X^{\text{gb}}-1)$ are the sampling intervals along the $z$- and $x$-axes, respectively. Accordingly, the codebook can be expressed as
\begin{equation}
\mathbf{U}_{\mathrm{T}}^{\mathrm{D}}
=
\left[
\mathbf{a}(\bar{z}_1,\bar{x}_1),\,
\dots,\,
\mathbf{a}(\bar{z}_{G_Z^{\text{gb}}},\bar{x}_{G_X^{\text{gb}}})
\right]^H
\in \mathbb{C}^{{|\mathcal{G}_{\text{gb}}|} \times N},
\label{eq:global_codebook}
\end{equation}
where $\mathbf{a}(\bar{z}_i,\bar{x}_j)\in\mathbb{C}^{N\times 1}$ is the near-field steering vector defined in \eqref{eq:steering_vector}.
The magnitude of the projected signal $\mathbf{y}_{\mathrm{C}}^{\mathrm{D}}(t)$ is first normalized and scaled by:
\begin{equation}
\mathbf{y}_{\text{norm}}^{\text{D}}(t)
=
\frac{
\big| \mathbf{y}_{\text{C}}^{\text{D}}(t) \big|
- \big| \mathbf{y}_{\text{C}}^{\text{D}}(t) \big|_{\text{min}}
}{
\big| \mathbf{y}_{\text{C}}^{\text{D}}(t) \big|_{\text{max}}
-\big| \mathbf{y}_{\text{C}}^{\text{D}}(t) \big|_{\text{min}}
}
\times 255.
\label{eq:normalization}
\end{equation}
The Cartesian-domain projection and the normalization are used only for signal-image generation and preprocessing, rather than as feasibility-enforcing projection steps in constrained optimization.
By reshaping $\mathbf{y}_{\text{norm}}^{\text{D}}(t)$ into an $G_Z \times G_X$ matrix, a single-channel grayscale signal image representation is obtained.} {\bl For visual clarity here, this normalized pixel value is duplicated across RGB channels to construct an RGB image, as shown in Fig.~\ref{fig:channel_images_rgb}(a). This image represents the energy distribution of the received signal in the transformed Cartesian domain. It can be seen that, each propagation path forms a clear X-shaped energy convergence pattern, whose central intersection (keypoint) indicates the location of the path center, i.e., the UE or a scatterer, as mathematically proven in~\cite{keypointCE}. These images serve as the basis for the subsequent detection and tracking stages.} However, achieving this ``Optimal'' image quality requires $N$ RF chains, imposing prohibitive hardware cost and power consumption in XL-MIMO systems.

To generate high-quality Cartesian-domain signal images from the received signals under HBF architectures with limited RF chains, we introduce an effective HBF codebook $\mathbf{U}_{\mathrm{T}}^{\mathrm{HBF}}$ and perform the projection by
\begin{equation}
\mathbf{y}_{\mathrm{C}}^{\mathrm{HBF}}(t)
=
\mathbf{U}_{\mathrm{T}}^{\mathrm{HBF}} \mathbf{y}(t),
\label{eq:proj}
\end{equation}
where $\mathbf{U}_{\mathrm{T}}^{\mathrm{HBF}} = \mathbf{U}_{\mathrm{T}}^{\mathrm{D}}\mathbf{W}^{\dagger}$ and $\mathbf{U}_{\mathrm{T}}^{\mathrm{HBF}} \in \mathbb{C}^{|\mathcal{G}_{\text{gb}}| \times N_{\mathrm{RF}}}$. By defining the projection matrix $\mathbf{P} = \mathbf{W}^{\dagger}\mathbf{W}$, the Cartesian-domain receive signal can be rewritten as
{\bl
\begin{equation}
\mathbf{y}_{\mathrm{C}}^{\mathrm{HBF}}(t)
=
\sqrt{P_r}\,\mathbf{U}_{\mathrm{T}}^{\mathrm{D}} \mathbf{P}\mathbf{h}(t)
+\mathbf{U}_{\mathrm{T}}^{\mathrm{HBF}}\tilde{\mathbf n}(t).
\end{equation}}
Following the same normalization and reshaping procedure defined in \eqref{eq:normalization}, we obtain the signal image $\mathbf{y}_{\text{img}}^{\text{HBF}}(t)$ under HBF architectures, as shown in Fig.~\ref{fig:channel_images_rgb}(b-h), where the resulting spatial energy distribution is determined by the design of $\mathbf{W}$. Specifically, the analog combining matrix defines a low-dimensional spatial measurement operator that captures the channel $\mathbf{h}(t)$ only through the row space spanned by $\mathbf{W}$. As a result, channel components that are not sufficiently observable under the current combiner cannot be recovered from the compressed observation alone by subsequent digital processing with the Cartesian codebook $\mathbf{U}_{\mathrm{T}}^{\mathrm{D}}$. If the row space of $\mathbf{W}$ is not well aligned with the physical spatial structure of the dominant propagation paths, the resulting measurement distortion may significantly degrade the energy distribution in the generated image $\mathbf{y}_{\text{img}}^{\text{HBF}}(t)$. Therefore, to ensure high-quality signal images, the design of $\mathbf{W}$ should satisfy two key requirements: 1) a row-orthogonal or low-correlation structure for stable sensing and improved numerical robustness, and 2) sufficient spatial coverage of the observation region to preserve the dominant propagation features.

Motivated by these requirements, constructing $\mathbf{W}$ using row-orthogonal discrete Fourier transform (DFT) beams is a natural choice under the fully connected (FC) HBF architecture (\textit{Baseline 2}). However, when the RF-chain budget is limited, simply selecting a subset of $N_{\mathrm{RF}}$ rows from the $N$-point DFT matrix leads to severe angular blind zones. As illustrated in Fig.~\ref{fig:channel_images_rgb}(b) and (c), paths within the covered angular region can still preserve nearly fully-digital-like clarity, whereas paths falling into the blind region become highly indistinguishable, causing severe loss of critical path information.

\subsection{Proposed Subarray-Based TM-PC HBF Scheme}
\label{subsec:proposed_hbf_scheme}

To overcome the bottleneck between hardware overhead and imaging quality, we propose a subarray-based time-multiplexed partially-connected (TM-PC) HBF scheme.

\subsubsection{Hardware Architecture: Partially Connected (PC) Topology}
\begin{figure}[t]
\centering
\includegraphics[width=0.9\linewidth]{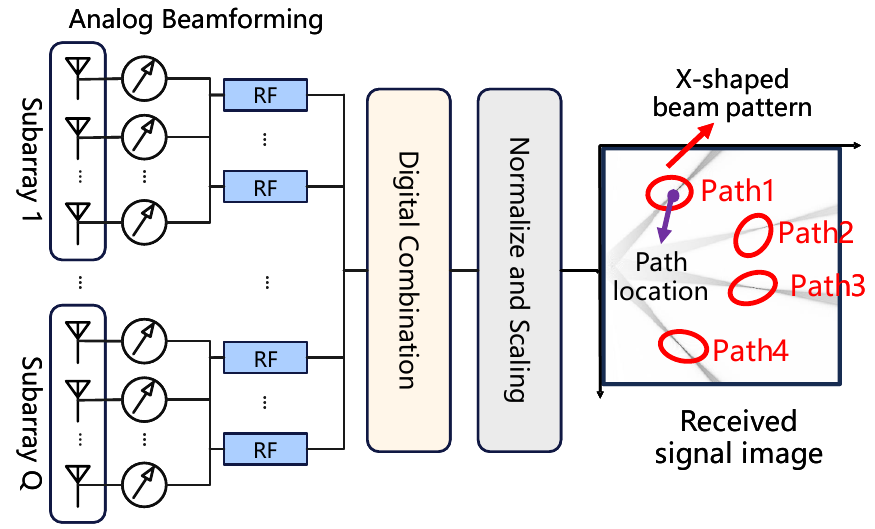}
\caption{{\bl The architecture of the proposed TM-PC HBF for signal image generation, comprising analog beamforming, digital combining, and subsequent normalization and scaling.}}
\label{fig:HBF_architecture1}
\end{figure}

To alleviate the hardware burden, we adopt a subarray-based partially connected architecture. {\bl As illustrated in Fig.~\ref{fig:HBF_architecture1}, the $N$-element array is divided into $Q$ subarrays, with the RF chains accordingly divided and assigned to them. The proposed scheme activates a total of $N_{\mathrm{RF}}$ RF chains. Under the partially connected topology, each subarray contains $N/Q$ antennas and is connected with $N_{\mathrm{RF}}/Q$ RF chains.}
Under this architecture, the analog combining matrix $\mathbf{W}(t)\in\mathbb{C}^{N_{\mathrm{RF}}\times N}$ can be represented as:
\begin{equation}
\mathbf{W}(t)
=
\mathrm{blkdiag}\!\left(\mathbf{W}_1(t),\mathbf{W}_2(t),\dots,\mathbf{W}_Q(t)\right),
\label{eq:beamforming_matrix}
\end{equation}
where $\mathbf{W}_q(t)\in\mathbb{C}^{(N_{\mathrm{RF}}/{Q})\times (N/Q)}$ denotes the combining matrix associated with the $q$-th subarray. Accordingly, the channel is partitioned as
\begin{equation}
\mathbf{h}(t)=\left[\mathbf{h}_1^{\top}(t),\dots,\mathbf{h}_Q^{\top}(t)\right]^{\top},
\end{equation}
{\bl where $\mathbf{h}_q(t)\in\mathbb{C}^{(N/Q)\times 1}$ denotes the sub-channel corresponding to the $q$-th subarray.}
Compared with the FC HBF architecture, this partially connected topology reduces the number of required phase shifters from $N N_{\mathrm{RF}}$ to $N_{\mathrm{RF}}^2$, thereby reducing hardware and power consumption.

\subsubsection{Proposed Reference Case: PC HBF with $N_{\mathrm{RF}} = N/2$}

We first establish a reference case to demonstrate that the PC HBF architecture can generate signal images with complete angular coverage. In this configuration, even when more RF chains are available, activating only $N_{\mathrm{RF}} = N/2$ chains is sufficient to maintain high-fidelity imaging. We divide the antenna array into $Q = 2$ subarrays and employ a complementary edge-split of the DFT matrix as follows.

\begin{figure*}[t]
\centering
\includegraphics[width=0.75\linewidth]
{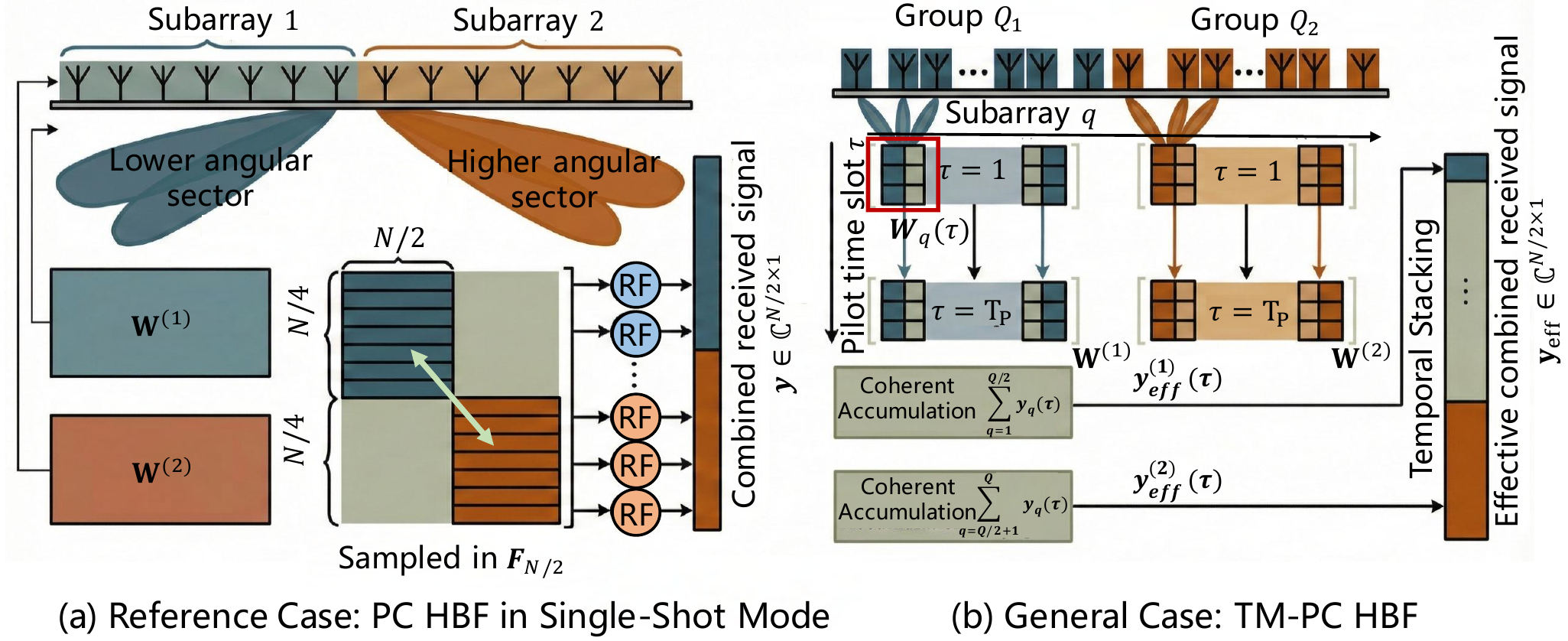}
\caption{The detailed design of the proposed TM-PC HBF framework, including the analog beamforming and time-multiplexed (TM) digital signal combining. (a) Reference Case ($N_{\mathrm{RF}} = N/2$): Achieves full spatial coverage in a single time slot by utilizing spatially complementary subarrays. (b) General Case ($N_{\mathrm{RF}} < N/2$): Synthesizes signals over $T_P$ pilot slots via TM to accumulate full spatial coverage under limited RF chains.}
\label{fig:HBF_architecture}
\end{figure*}

As shown in Fig.~\ref{fig:HBF_architecture}(a), the analog combiner is constructed by partitioning the $N/2$-point DFT matrix $\mathbf{F}_{N/2}$, into two complementary sub-matrices: $\mathbf{W}^{(1)}$ comprising the first $N/4$ rows, and $\mathbf{W}^{(2)}$ containing the remaining $N/4$ rows. The resulting reference combiner is:
\begin{equation}
\mathbf{W}_{\mathrm{ref}} = \mathrm{blkdiag} \left( \mathbf{W}^{(1)}, \mathbf{W}^{(2)} \right).
\end{equation}
The row spaces of $\mathbf{W}^{(1)}$ and $\mathbf{W}^{(2)}$ are complementary, jointly spanning the entire discrete DFT beamspace for the two-subarray architecture. Consequently, unlike the FC HBF design in Baseline~2, this design eliminates the angular blind region caused by incomplete row selection. The received signal is thus given by
\begin{equation}
\mathbf{y}_{\mathrm{ref}}(t) = \sqrt{P_r}\,\mathbf{W}_{\mathrm{ref}}\big(\mathbf{h}(t)+\mathbf{n}(t)\big).
\end{equation}
By applying the spatial projection \eqref{eq:proj} along with the normalization and scaling \eqref{eq:normalization}, the received signal $\mathbf{y}_{\mathrm{ref}}(t)$ is converted into a signal image, as shown in Fig.~\ref{fig:channel_images_rgb}(d). It preserves a clear and complete energy convergence region featuring highly distinguishable keypoints.

{\bl
\subsubsection{Proposed General Case: TM-PC HBF with $N_{\mathrm{RF}} < N/2$}

When the available RF chains are less than $N/2$, a straightforward approach in this regime is to divide the array into $Q = N/N_{\mathrm{RF}}>2$ subarrays and assign adjacent DFT rows to each subarray (\textit{Baseline 3}: the non-multiplexed PC HBF). This design maintains low hardware overhead, since it uses only $N_{\mathrm{RF}}$ RF chains and $N_{\mathrm{RF}}^2$ phase shifters. However, the reduced subarray aperture severely degrades the signal image quality. As shown in Fig.~\ref{fig:channel_images_rgb}(e), when the number of RF chains is small, the reduced subarray aperture causes the energy convergence region to spread out, making the path center much less distinguishable.

\begin{figure*}[t]
    \centering
        \includegraphics[width=0.99\linewidth]{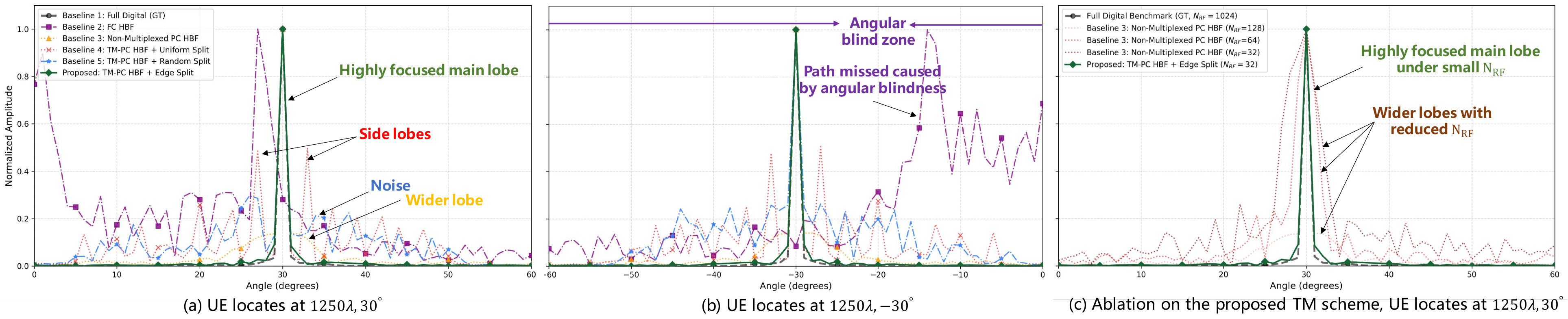} 
        \caption{Comparison of the angular spectrum distributions of the generated signal image under different HBF schemes with $N_{\text{RF}}=N/8$ for a representative UE located at a distance of $1024\lambda$ from the ULA center. The spatial focusing capabilities are evaluated at target angles of (a) $\phi_0=30^\circ$ and (b) $\phi_0=-30^\circ$. {\bl Furthermore, (c) presents an ablation study and comparison of the proposed TM HBF mechanism at $\phi_0=30^\circ$.}}
    \label{fig:angular_spectrum}
\end{figure*}

\begin{table*}[t]
\centering
\begin{threeparttable}
\caption{{\bl Comparative analysis of the system overhead and generated signal image quality under different HBF schemes.}}
\label{tab:overhead_comparison}
\renewcommand{\arraystretch}{1.25}
\setlength{\tabcolsep}{4pt}
\small
\begin{tabular}{@{}lccccc@{}}
\toprule
\textbf{Scheme} & \textbf{RF Chains} & \textbf{Phase Shifters} & \textbf{TM Slots ($T_P$)} & \textbf{Image Quality} & \textbf{Hardware Cost} \\
\midrule
Baseline 1: Full Digital & $N$ & $0$ & $1$ & Optimal & High \\

Baseline 2: FC HBF & $N_{\mathrm{RF}}$ & $N N_{\mathrm{RF}}$ & $1$ & Angular Blind Region & High \\

Baseline 3: Non-TM-PC HBF & $N_{\mathrm{RF}}$ & $N_{\mathrm{RF}}^2$ & $1$ & Narrow Convergence Zone & Low \\

Baseline 4: TM-PC HBF with Random Split & $N_{\mathrm{RF}}$ & $N_{\mathrm{RF}}^2$ & $NQ/(2N_{\mathrm{RF}})$ &Noise Interference & Low \\

Baseline 5: TM-PC HBF with Uniform Split & $N_{\mathrm{RF}}$ & $N_{\mathrm{RF}}^2$ & $NQ/(2N_{\mathrm{RF}})$ & Sidelobe Interference & Low \\
\midrule
\rowcolor{gray!15} \textbf{Reference Case: PC HBF ($N_{\mathrm{RF}}=N/2$)} & $N/2$ & $N^2/4$ & $1$ & \textbf{Clear, Complete Region} & \textbf{Low} \\

\rowcolor{gray!15} \textbf{Proposed TM-PC HBF ($N_{\mathrm{RF}}<N/2$)} & $N_{\mathrm{RF}}$ & $NN_{\mathrm{RF}}/Q^2$ & $NQ/(2N_{\mathrm{RF}})$ & \textbf{Same as Reference} & \textbf{Low} \\
\bottomrule
\end{tabular}
\end{threeparttable}
\end{table*}

To overcome this limitation, we exploit the \textit{temporal dimension} through a time-multiplexed (TM) mechanism.\footnote{{\bl The TM mechanism preserves the original partially connected topology and only reconfigures the analog coefficients across pilot sub-slots, while the effective reference observation is synthesized digitally.}}
Let $t$ denote the index of the main time slot for channel tracking, and $\tau \in \{1,\dots,T_P\}$ denote the pilot sub-slot index within the main time slot. The key idea is that, while only $N_{\mathrm{RF}}$ RF chains are available during each pilot sub-slot, the signals collected over the $T_P$ sub-slots can be temporally aggregated, as shown in Fig.~\ref{fig:HBF_architecture}(b). Through proper design of the sub-slot combiners, this aggregated signal is equivalent to the received signal $\mathbf{y}_{\mathrm{ref}}(t)$ from the reference case ($N_{\mathrm{RF}}=N/2$), ensuring the generation of the same high-quality signal image.

Specifically, the number of pilot sub-slots is set as $T_P = NQ/(2N_{\mathrm{RF}})$, which serves as the multiplexing factor.
More generally, given $N_{\mathrm{RF}}^{\mathrm{ava}}$ physically available RF chains, the proposed scheme selects the minimum feasible multiplexing factor, given by
\begin{equation}
T_P=\left\lceil \frac{NQ}{2N_{\mathrm{RF}}^{\mathrm{ava}}}\right\rceil,
\end{equation}
and the number of activated RF chains is thus determined as
\begin{equation}
N_{\mathrm{RF}}=\frac{NQ}{2T_P}\leq N_{\mathrm{RF}}^{\mathrm{ava}}.
\end{equation}
Assuming the channel remains quasi-static over the $T_P$ pilot sub-slots, the temporal aggregation is valid. This intra-slot pilot configuration aligns with practical uplink sounding procedures in multicarrier systems, e.g., SRS bursts over a subset of OFDM symbols within one slot~\cite{3GPP3811,3GPP3813}, and is applicable when the total duration of the pilot sub-slots is within the channel coherence time.

To synthesize a received signal identical to that of the reference case, the $Q$ subarrays are logically partitioned into two disjoint groups defined as
\begin{equation}
\mathcal{Q}_1 = \left\{1,\dots,Q/2\right\}, \qquad
\mathcal{Q}_2 = \left\{Q/2+1,\dots,Q\right\}.
\end{equation}
The subarrays in $\mathcal{Q}_1$ collectively synthesize the rows of $\mathbf{W}^{(1)}$, while those in $\mathcal{Q}_2$ synthesize the rows of $\mathbf{W}^{(2)}$.
Accordingly, at the $\tau$-th pilot sub-slot of the $t$-th main time slot, the analog coefficient block associated with the $q$-th subarray is selected as
\begin{equation}
\mathbf{W}_{q}(t,\tau) =
\begin{cases}
\mathbf{W}^{(1)} \left[ \mathcal{I}_{\tau}, \mathcal{J}_{\bar q}^{\rm loc} \right], & q \in \mathcal{Q}_1, \\
\mathbf{W}^{(2)} \left[ \mathcal{I}_{\tau}, \mathcal{J}_{\bar q}^{\rm loc} \right], & q \in \mathcal{Q}_2,
\end{cases}
\label{eq:subarray_combiner}
\end{equation}
where $\mathbf{W}_{q}(t,\tau)\in\mathbb{C}^{({N_{\mathrm{RF}}}/{Q})\times ({N}/{Q})}$,
$\bar q=q$ for $q\in\mathcal{Q}_1$ and $\bar q=q-Q/2$ for $q\in\mathcal{Q}_2$.
The row and local column index sets are defined as
\begin{equation}
\mathcal{I}_{\tau}
=
\left[
\tfrac{(\tau-1)N_{\text{RF}}}{2Q} :
\tfrac{\tau N_{\text{RF}}}{2Q}-1
\right],
\quad
\mathcal{J}_{\bar q}^{\rm loc}
=
\left[
\tfrac{(\bar q-1)N}{Q} :
\tfrac{\bar q N}{Q}-1
\right].
\label{eq:index_sets}
\end{equation}

The contribution from the $q$-th subarray to the group-wise effective observation can be written as
\begin{equation}
\mathbf{y}_q(t,\tau)
=
\sqrt{P_r}\,\mathbf{W}_q(t,\tau)\mathbf{h}_q(t)+\mathbf{n}_q(t,\tau),
\end{equation}
where $\mathbf{y}_q(t,\tau)\in\mathbb{C}^{({N_{\mathrm{RF}}}/{2Q})\times 1}$,
and $\mathbf{h}_q(t)\in\mathbb{C}^{N/Q\times 1}$ denotes the channel vector associated with the $q$-th subarray, and $\mathbf{n}_q(t,\tau)\in\mathbb{C}^{({N_{\mathrm{RF}}}/{2Q})\times 1}$ is the corresponding noise contribution.
At each pilot sub-slot $\tau$, the contributions from the subarrays in the same group are aggregated as
\begin{equation}
\mathbf{y}_{\mathrm{eff}}^{1}(t,\tau) = \sum_{q \in \mathcal{Q}_1}\mathbf{y}_q(t,\tau), \quad
\mathbf{y}_{\mathrm{eff}}^{2}(t,\tau) = \sum_{q \in \mathcal{Q}_2}\mathbf{y}_q(t,\tau).
\end{equation}

Then, by concatenating these group-wise aggregated observations across all $T_P$ pilot sub-slots, the aggregated received signal is obtained as
\begin{equation}
\begin{split}
\mathbf{y}_{\mathrm{eff}}(t)
= \Big[
& \mathbf{y}_{\mathrm{eff}}^{1}(t,1)^{\top}, \dots, \mathbf{y}_{\mathrm{eff}}^{1}(t,T_P)^{\top}, \\
& \mathbf{y}_{\mathrm{eff}}^{2}(t,1)^{\top}, \dots, \mathbf{y}_{\mathrm{eff}}^{2}(t,T_P)^{\top}
\Big]^{\top}
\in \mathbb{C}^{\frac{N}{2}\times 1}.
\end{split}
\end{equation}
Under the intra-slot quasi-static assumption, $\mathbf{y}_{\mathrm{eff}}(t)$ can be equivalently written as a full-aperture observation:
\begin{equation}
    \mathbf{y}_{\mathrm{eff}}(t) = \mathbf{y}_{\mathrm{ref}}(t)
    =\sqrt{P_r}\,\mathbf{W}_{\mathrm{eff}} \mathbf{h}(t) + \tilde{\mathbf{n}}_{\mathrm{eff}}(t),
    \label{eq:eff_observation}
\end{equation}
where
\begin{equation}
\mathbf{W}_{\mathrm{eff}} = \mathbf{W}_{\mathrm{ref}}
= \mathrm{blkdiag}\!\left(\mathbf{W}^{(1)}, \mathbf{W}^{(2)}\right)
\in \mathbb{C}^{\frac{N}{2}\times N},
\end{equation}
and $\tilde{\mathbf{n}}_{\mathrm{eff}}(t) \in \mathbb{C}^{\frac{N}{2}\times 1}$ denotes the aggregated effective noise.
It should be noted that the TM acquisition is performed within each main tracking slot.
Unless otherwise specified, the notation $\mathbf{y}(t)$ and $\mathbf{W}$ used in the subsequent OMP-based refinement and tracking modules refers to
\begin{equation}
\mathbf{y}(t)\triangleq \mathbf{y}_{\mathrm{eff}}(t), 
\qquad 
\mathbf{W}\triangleq \mathbf{W}_{\mathrm{eff}}.
\end{equation}
}

We further compare the proposed TM-PC HBF with two DFT split schemes for $\mathbf{W}^{(1)}$ and $\mathbf{W}^{(2)}$ construction, i.e., random split and uniform split. As shown in Fig.~\ref{fig:channel_images_rgb}(f), the random-split design introduces pronounced noise interference, which makes the path-center locations difficult to distinguish. As shown in Fig.~\ref{fig:channel_images_rgb}(g), the uniform-split design suffers from relatively strong sidelobe energy. When multiple paths coexist, the sidelobe responses overlap and lead to noticeable sidelobe interference. By comparison, the proposed edge-split design can preserve the clear convergence zone and enable more accurate localization of the path centers.

\subsection{Angular-Spectrum Analysis}
To further explain the signal-image characteristics of different HBF schemes, we examine the angular spectrum in Fig.~\ref{fig:angular_spectrum}. 
For representative paths at $30^\circ$ and $-30^\circ$, the proposed TM-PC HBF consistently produces a sharp dominant peak around the ground truth (GT) angle, with a beamwidth close to that of the full-digital benchmark, indicating high angular resolution under limited RF resources. 
In contrast, the FC-HBF baseline suffers from angular blind regions, the non-multiplexed PC-HBF baseline exhibits increasingly wider lobes as $N_{\mathrm{RF}}$ is reduced due to aperture loss, and alternative TM-PC codebook designs introduce stronger sidelobes or noise fluctuations. 
Moreover, the proposed time-multiplexed mechanism effectively compensates for the aperture loss caused by using fewer RF chains by aggregating complementary subarray observations across multiple sub-slots, thereby maintaining a concentrated main lobe and preserving angular focusing capability even at low RF-chain budgets.

As a summary, Table~\ref{tab:overhead_comparison} provides a detailed comparison of the hardware overhead and received image quality under different beamforming schemes. Compared to the baselines, our design strikes an optimal balance: it substantially reduces hardware overhead via the PC HBF architecture, while achieving high image quality through the TM scheme.

\section{Learning-Assisted Fast Detection and Tracking} 
\label{sec: Learning-Assisted Fast Detection and Tracking}
Building on the generated signal images, we next develop a learning-assisted framework for efficient trajectory and channel tracking.

\begin{figure*}[t]
\centering
\includegraphics[width=0.85\linewidth]{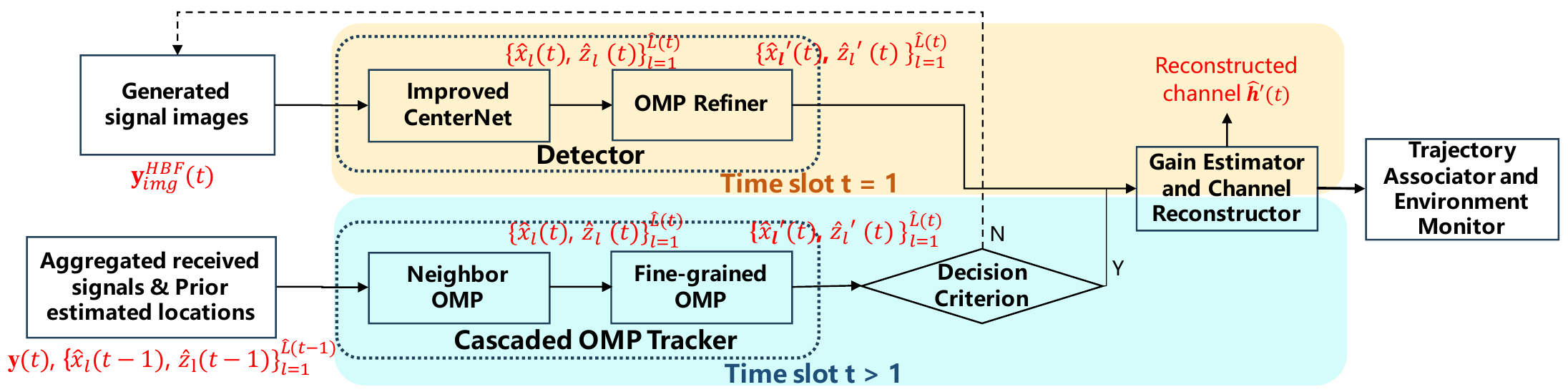}
\caption{{\bl Overall framework of the proposed efficient joint trajectory and channel tracking algorithm.}}
\label{fig:algorithm_framework}
\end{figure*}

{\bl
\subsection{Overall Workflow}
As illustrated in Fig.~\ref{fig:algorithm_framework}, the proposed framework operates sequentially. 
At each main slot $t$, the TM-PC HBF provides the effective received signal $\mathbf{y}_{\mathrm{eff}}(t)$, which is treated as $\mathbf{y}(t)$ in the OMP-based modules. 
At $t=1$, $\mathbf{y}_{\mathrm{eff}}(1)$ is converted into a Cartesian-domain signal image for CenterNet-based path initialization, followed by local OMP refinement and LS gain estimation. 
For $t>1$, the cascaded OMP tracker directly updates the path locations from $\mathbf{y}_{\mathrm{eff}}(t)$ using the previous-slot estimates as priors, without repeatedly generating signal images. 
The signal image and detector are re-activated only when the residual-based validity check fails.
}

\subsection{Improved CenterNet for Fast Path Detection}

{\bl The path position estimation problem can be reformulated as a point-based object detection problem on the generated signal image. Different from conventional object detection tasks in computer vision, our goal is not to predict the width and height of the target region, but rather to accurately localize the center positions of the ``X-shaped'' path. Therefore, the considered task is more closely related to precise center-point detection, for which mature point-based detection frameworks have been extensively developed in computer vision~\cite{centernet,cornernet,objectsaspoints,hrnetpose,centertrack,centerpoint}. Based on this insight, we develop an improved CenterNet-based framework~\cite{centernet}, dedicated for accurate center localization of propagation paths in the signal image.}

\begin{figure*}[t]
\centering
\includegraphics[width=0.9\linewidth]{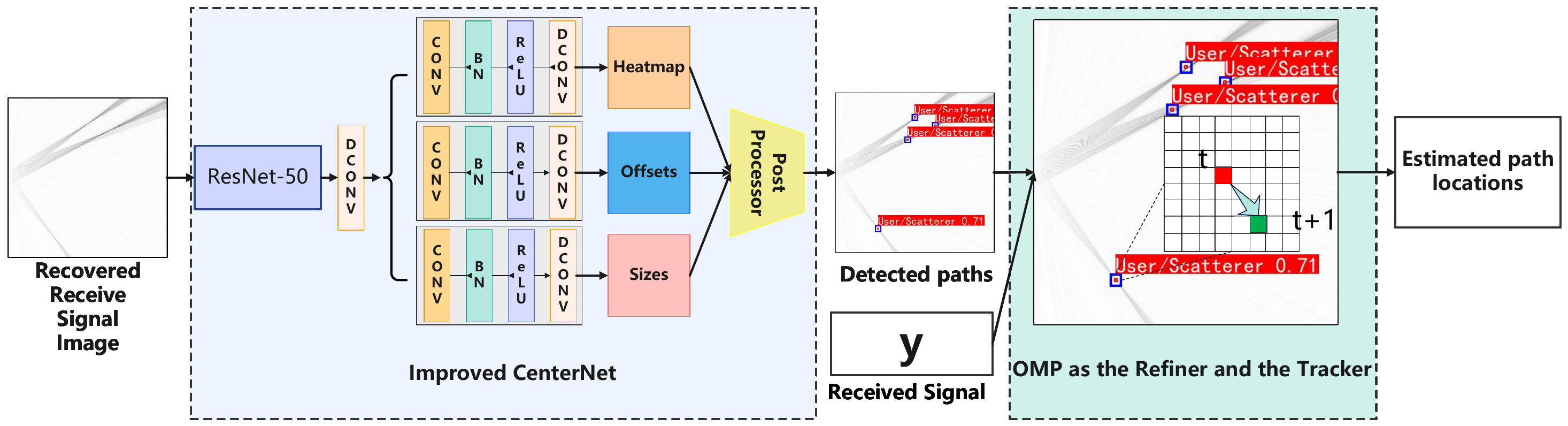}
\caption{Architecture of the improved CenterNet network, followed by the OMP-based refiner and cascaded OMP tracker.}
\label{fig:centernet}
\end{figure*}

The architecture of the improved CenterNet is illustrated in Fig.~\ref{fig:centernet}. 
Specifically, a ResNet-50 backbone is employed to extract shared features from the signal image $\mathbf{y}_{\text{img}}^{\text{HBF}}(t)$. 
The resulting feature maps are then fed into three parallel prediction heads, each consisting of a convolution (CONV) layer, batch normalization (BN), a rectified linear unit (ReLU), and a deconvolution (DCONV) layer. 
These heads produce three output maps of size $H_O \times W_O$, corresponding to: 1) a Gaussian heatmap for propagation-path keypoint detection; 2) a size map characterizing the spatial extent of each detected region; and 3) an offset map for sub-pixel localization refinement. 
Finally, a post-processing module removes low-confidence candidates and applies non-maximum suppression (NMS) to eliminate redundant detections.

{\bl Compared with our prior image-based keypoint detector in~\cite{keypointCE}, the proposed CenterNet introduces three essential upgrades for dynamic tracking. First, by replacing direct coordinate regression with heatmap-based detection, it can accommodate the birth-death behavior of mobile scatterers, i.e., a time-varying $L(t)$. Second, to emphasize accurate center localization, we adopt a unified detection scale and use the wing loss~\cite{wingloss}, which emphasizes small localization errors and provides more accurate coordinate priors for the subsequent local OMP refinement. Third, the detector is no longer used as an isolated static estimator. Instead, it is integrated into the tracking loop as a fast initializer and re-activation module, working with the residual-based validity check of the cascaded OMP tracker to recover from severe drift and detect newly appeared scatterers.}

{\bl It is worth noting that the improved CenterNet is introduced only for the intermediate sub-task of coarse path-center detection on the generated signal images, rather than as an end-to-end optimizer of the final communication metric. Therefore, the following loss is adopted as a task-specific training objective for accurate keypoint localization.}
The overall loss function for training is defined as
\begin{equation}
\mathcal{L} = \lambda_{\text{hm}}L_{\text{heatmap}}+\lambda_{\text{off}}L_{\text{offset}}+\lambda_{\text{size}}L_{\text{size}},
\label{eq:loss}
\end{equation}
where $\lambda_{\text{hm}}$, $\lambda_{\text{off}}$, and $\lambda_{\text{size}}$ denote the respective weights. Since our primary objective is accurate keypoint localization of propagation paths rather than precise estimation of the width and height of the corresponding ``X-shaped'' region, we assign a relatively smaller weight to $L_{\text{size}}$, such that the optimization focuses more on heatmap prediction and offset refinement. The heatmap loss is a cross-entropy loss defined as 
\begin{equation}
\small
L_{\text{heatmap}} = -\frac{1}{N} \sum_{x, y, c} \begin{cases} 
(1 - \hat{Y}_{xyc})^\alpha \log(\hat{Y}_{xyc}), \quad Y_{xyc}=1 \\
(1 - Y_{xyc})^\beta (\hat{Y}_{xyc})^\alpha \log(1 - \hat{Y}_{xyc}), \quad \text{else,}
\end{cases}
\label{eq:heatmap_loss}
\end{equation}
where $\alpha$ and $\beta$ are hyperparameters regulating the penalty reduction. $Y_{xyc} \in \{0,1\}$ is the GT label indicating the presence (1) or absence (0) of a propagation path at location $(x, y)$, with $\hat{Y}_{xyc}$ denoting the predicted probability. The offset and size losses are defined using the wing loss as
\begin{equation}
L_{\text{wing}}(d)= \begin{cases}\omega \ln (1+|d| / \epsilon), & \text { if }|d|<\omega, \\ |d|-C, & \text {otherwise,}\end{cases}
\end{equation}
where $\omega$ and $\epsilon$ are hyperparameters of wing loss, $d$ represents the prediction errors, including $d=p-\hat{p}$ for offset or $d=s-\hat{s}$ for size prediction, and the constant $C = \omega - \omega \ln(1 + \omega / \epsilon)$.

Following the network output, a post-processing module is used to refine the detections. First, candidates with confidence scores below a preset threshold $C_{\text{thre}}$ are removed. Then, a non-maximum suppression (NMS) procedure is applied to suppress redundant neighboring detections~\cite{nms}.

{\bl
\subsection{OMP Refiner and Channel Estimator}

Finally, as shown in Fig.~\ref{fig:centernet}, an orthogonal matching pursuit (OMP) refiner is employed to perform fine-grained matching around the detected keypoint locations, thereby further improving localization accuracy. Specifically, using the coarse coordinates predicted by the improved CenterNet as spatial priors, we construct a local refinement grid $\mathcal{G}_{\mathrm{ref}}=\{(z_i,x_i)\}_{i=1}^{G_Z^{\mathrm{ref}}G_X^{\mathrm{ref}}}$ with sampling interval $(\Delta_Z^{\text{ref}},\Delta_X^{\text{ref}})$ over the neighborhoods of the detected locations, whose cardinality is
\begin{equation}
|\mathcal{G}_{\mathrm{ref}}|=G_Z^{\mathrm{ref}}G_X^{\mathrm{ref}}.
\end{equation}
Accordingly, the corresponding local dictionary is constructed as:
\begin{equation}
\begin{aligned}
\mathbf{\Psi}_{\text{ref}} 
&= [\boldsymbol{\psi}_1,\boldsymbol{\psi}_2,\ldots,\boldsymbol{\psi}_{G_{\text{ref}}}] \\
&= [\mathbf{a}(z_1,x_1),\mathbf{a}(z_2,x_2),\ldots,\mathbf{a}(z_{G_{\text{ref}}},x_{G_{\text{ref}}})],
\end{aligned}
\end{equation}
where each atom $\boldsymbol{\psi}_i=\mathbf{a}(z_i,x_i)\in\mathbb{C}^{N\times 1}$ is the steering vector evaluated at the $g$-th local grid point according to~\eqref{eq:steering_vector}. The OMP refiner then solves the following $\ell_0$-constrained sparse recovery problem:
\begin{equation}
    \min_{\mathbf{c}} \left\| \mathbf{y} - \mathbf{W} \mathbf{\Psi}_{\text{ref}} \mathbf{c} \right\|_2^2
    \quad \text{s.t.} \quad \|\mathbf{c}\|_0 \le \hat{L},
\end{equation}
where $\mathbf{y}$ denotes the received signal, $\mathbf{W}$ is the combining matrix, $\mathbf{c} \in \mathbb{C}^{G_{\text{ref}} \times 1}$ is the sparse coefficient vector, and $\hat{L}$ is the number of valid propagation paths detected by the improved CenterNet. By iteratively selecting the dictionary atoms most correlated with the residual signal, OMP identifies the support of $\mathbf{c}$ and obtains the refined path coordinates $\{(\hat{z}_l,\hat{x}_l)\}_{l=1}^{\hat{L}}$. Based on these refined coordinates, we form the full-dimensional steering matrix as
\begin{equation}
    \mathbf{A}_{\text{LS}} = [\mathbf{a}(\hat{z}_1,\hat{x}_1), \mathbf{a}(\hat{z}_2,\hat{x}_2), \ldots, \mathbf{a}(\hat{z}_{\hat{L}},\hat{x}_{\hat{L}})] \in \mathbb{C}^{N \times \hat{L}}.
\label{eq:form_steer}
\end{equation}
The effective sensing matrix is defined as $\mathbf{A}_{\text{LS}}^{\text{eff}}=\mathbf{W}\mathbf{A}_{\text{LS}}$. Then, the complex path gains are estimated via LS as
\begin{equation}
    \hat{\boldsymbol{G}} =
    \left( (\mathbf{A}_{\text{LS}}^{\text{eff}})^H \mathbf{A}_{\text{LS}}^{\text{eff}} \right)^{-1}
    (\mathbf{A}_{\text{LS}}^{\text{eff}})^H \mathbf{y}.
\label{eq:ls_gain_estimation}
\end{equation}
Finally, the channel can be estimated by
\begin{equation}
    \hat{\mathbf{h}}
    = \mathbf{A}_{\text{LS}} \hat{\boldsymbol{G}}.
\label{eq:recon_h}
\end{equation}
Since the refinement is carried out on a finite local dictionary and the sparsity level is bounded by the detected path number, the OMP procedure terminates after a finite number of selections. Moreover, given the refined path support, the subsequent LS update provides the closed-form least-squares estimate of the corresponding path gains.
}

\subsection{User, Scatterers, and Channel Tracker}
We further exploit the geometric information of the mobile user and scatterers to perform efficient tracking.

\begin{algorithm}[t]
\caption{{\bl Cascaded OMP Tracker.}}
\label{alg:cascaded_omp}
\begin{algorithmic}[1]
\REQUIRE Received signal $\mathbf{y}(t)$; prior estimated path set $\{(\hat{z}_l(t\!-\!1),\hat{x}_l(t\!-\!1))\}_{l=1}^{\hat{L}(t-1)}$; thresholds $\epsilon_{\mathrm{th}}$ and $P_{\mathrm{fa}}$.
\ENSURE Updated path set $\{(\hat{z}_l(t),\hat{x}_l(t))\}_{l=1}^{\hat{L}(t)}$, channel estimate $\hat{\mathbf{h}}(t)$, tracking status.

\FOR{$l=1$ \TO $\hat{L}(t-1)$}
    \STATE Perform Neighbor OMP in a coarse local window centered at $(\hat{z}_l(t\!-\!1),\hat{x}_l(t\!-\!1))$ to obtain an intermediate coordinate.
    \STATE Perform fine-grained OMP in a finer local window around the intermediate coordinate to obtain $(\hat{z}_l(t),\hat{x}_l(t))$.
\ENDFOR
\STATE Estimate path gains via \eqref{eq:ls_gain_estimation} and reconstruct channel $\hat{\mathbf{h}}(t)$ via \eqref{eq:recon_h}.
\STATE Compute residual signal $\mathbf{y}_{\mathrm{res}}(t)$ as~\eqref{eq:y_res} and relative ratio $\eta(t) = \|\mathbf{y}_{\mathrm{res}}(t)\|_2^2 / \|\mathbf{y}(t)\|_2^2$.

\IF{$\eta(t) < \epsilon_{\mathrm{th}}$}
    \STATE \COMMENT{\textit{Primary Criterion: Residual energy is sufficiently low.}}
    \STATE Accept tracking result, set $\hat{L}(t)=\hat{L}(t-1)$.
\ELSIF{$T(t) < \gamma$ \COMMENT{\textit{Secondary Criterion: EVT-based statistical test \eqref{decision}.}}}
    \STATE Accept tracking result, set $\hat{L}(t)=\hat{L}(t-1)$, and return \textbf{Success}.
\ELSE
    \STATE \COMMENT{\textit{Validity check failed: Potential path birth or drift.}}
    \STATE Trigger improved CenterNet re-detection and refinement, and return \textbf{Re-detection}.
\ENDIF
\end{algorithmic}
\end{algorithm}

\subsubsection{Cascaded OMP for Tracking}
As illustrated in Fig.~\ref{fig:centernet}, the proposed two-stage strategy \textit{cascaded OMP} leverages the position estimates from the previous time slot to constrain the search region, thereby reducing computational complexity while maintaining high localization accuracy. The algorithmic procedure is given in Algorithm~\ref{alg:cascaded_omp}.

{\bl 
Specifically, for each tracked path at time slot $t$, a coarse local search window of size $G_Z^{\mathrm{nb}}\times G_X^{\mathrm{nb}}$ and sampling interval $(\Delta_Z^{\mathrm{nb}}\times \Delta_X^{\mathrm{nb}})$ is first centered around the previous-slot estimate, and Neighbor OMP is applied to obtain an intermediate coordinate. Then, a finer local refinement window of size $G_Z^{\mathrm{ref}}\times G_X^{\mathrm{ref}}$ and sampling interval $(\Delta_Z^{\mathrm{ref}}\times \Delta_X^{\mathrm{ref}})$ is constructed around this intermediate coordinate, within which fine-grained OMP is performed to obtain the final refined position.}

{\bl To robustly capture dynamic environmental changes, such as sudden path appearance or tracking failure, a \textit{hierarchical validity check} is applied at each time slot $t$.} The residual signal after canceling the contributions of the currently tracked paths can be expressed as:
\begin{equation}
\mathbf{y}_{\mathrm{res}}(t) = \mathbf{y}(t)-\mathbf{W}\hat{\mathbf{h}}(t)
\label{eq:y_res}
\end{equation}
Let $\eta(t) \triangleq \|\mathbf{y}_{\mathrm{res}}(t)\|_2^2 / \|\mathbf{y}(t)\|_2^2$ denote the relative residual ratio. If $\eta(t) < \epsilon_{\mathrm{th}}$, 
where $\epsilon_{\mathrm{th}}$ is a predefined tolerance threshold, the current tracked-path set is directly declared valid. This primary criterion avoids unnecessary re-detection caused by negligible residuals in high-SNR regimes.
In cases where $\eta(t) \ge \epsilon_{\mathrm{th}}$, a finer-grained statistical evaluation is performed to detect potential path birth or tracking divergence. Leveraging the statistical properties of high-dimensional noise vectors~\cite{CS_Tips_Tricks}, we adopt an extreme-value-theory (EVT)-based threshold and formulate the detection condition as
\begin{equation}
\begin{aligned}
T(t)\triangleq
\max_{(\bar{z},\bar{x})\in\mathcal{G}}
\frac{
\left|
\left(\mathbf{W}\mathbf{a}(\bar{z},\bar{x})\right)^H
\mathbf{y}_{\mathrm{res}}(t)
\right|^2
}{
\sigma^2
\left\|
\mathbf{W}\mathbf{a}(\bar{z},\bar{x})
\right\|_2^2
}
<
\gamma,
\end{aligned}
\label{decision}
\end{equation}
where $\gamma = \ln |\mathcal{G}_{\text{gb}}| - \ln\big(-\ln(1-P_{\mathrm{fa}})\big)$ is the EVT-derived threshold.

If the condition in~\eqref{decision} is violated, significant unmodeled energy is declared, indicating that the current tracked-path set is incomplete or unreliable. In such instances, the tracker abandons the current local update and re-performs path searching using the improved CenterNet and refinement to recover the missing or newly appeared paths. This dual-criteria mechanism ensures a robust balance between tracking continuity and detection sensitivity, effectively preventing permanent drift propagation while maintaining stability against high-SNR over-detection artifacts.

\subsubsection{Trajectory Association and Environment Monitoring}

Given the refined position estimates of all detected entities over $T$ time slots, we can further associate the per-slot detections into continuous trajectories. To this end, a Hungarian algorithm-based assignment method~\cite{Hungarian} is adopted. Specifically, for each pair of consecutive time slots $(t-1,t)$, a cost matrix is constructed between the detections at slot $t-1$ and those at slot $t$, where each entry is defined as the Euclidean distance between two estimated locations. The Hungarian algorithm is then applied to solve the resulting minimum-cost matching problem and obtain the optimal one-to-one assignments. Detections at the current slot that cannot be matched are treated as newly appeared paths, whereas previously tracked paths without valid matches are regarded as disappeared. Repeating this procedure for $t=2,\ldots,T$ links detections across time and yields the associated UE and scatterer trajectories.

{\bl These trajectory-level results provide the geometric basis for downstream environment monitoring. Specifically, the displacement statistics of each trajectory can be evaluated over time to distinguish static scatterers from dynamic objects, where trajectories with negligible displacement variation are classified as static, while those with sustained motion are regarded as dynamic. Building upon this classification, the estimated static features can be cross-checked against prior digital map information for map matching and sensing consistency validation. Furthermore, the appearance, disappearance, or abnormal motion of dynamic trajectories can be used to indicate environmental changes or anomalies.}

\section{Experimental Results} \label{sec:Experimental Results}

In this section, we first introduce the experimental setups and the baseline algorithms for comparison. We then compare the detection performance of the improved CenterNet under different HBF designs, followed by evaluations of the estimation performance at the initial time slot and the overall tracking performance across the entire observation window. Finally, the computational complexity is analyzed.


\begin{table}[t]
\centering
\caption{Experimental settings.}
\label{tab:simulation_params}
\renewcommand{\arraystretch}{1.1} 
\setlength{\tabcolsep}{8pt}      
\footnotesize 

\begin{tabular}{l c} 
\toprule
\textbf{Parameter} & \textbf{Value} \\
\midrule
\multicolumn{2}{l}{\textit{\textbf{1. System \& environment}}} \\
Central carrier frequency $f_c$ & $7$ GHz \\
Antenna array (ULA) & $N=1024$, $d=\lambda/2$ \\
Region bound of x-axis $[x_{\min}, x_{\max}]$ & $[-1280\lambda, 1280\lambda]$ \\
Region bound of z-axis $[z_{\min}, z_{\max}]$ & $[0, 2560\lambda]$ \\
Duration & $T=6$ slots ($0.5$ s/slot) \\
Speed range & $[0,5]$ m/s \\
Number of total paths & $[3,6]$ \\
Number of static paths $L_1$  & 2 \\
New/disappearance probability $(P_{\mathrm{n}},P_{\mathrm{d}})$ & $0.1,\,0.1$ \\

\midrule
\multicolumn{2}{l}{\textit{\textbf{2. Dataset \& training}}} \\
Train/valid/test set size & $20000/5000/4800$ \\
SNR range & $[-3,18]$ dB \\
Optimizer / Batch size & Adam / $128$ \\
Learning rate / momentum & $1 \times 10^{-3},\,0.9$ \\
Training epochs & $1000$ \\

\midrule
\multicolumn{2}{l}{\textit{\textbf{3. Improved CenterNet}}} \\
Input / Output resolution & $512 \times 512$ / $128 \times 128$ \\
Loss weights $(\lambda_{\mathrm{hm}},\lambda_{\mathrm{off}},\lambda_{\mathrm{size}})$ & $1,\,1,\,0.2$ \\
Focal loss $(\alpha,\beta)$ & $2,\,4$ \\
Wing loss $(\omega,\epsilon)$ & $10,\,2$ \\
Confidence threshold $C_{\text{thre}}$ & $0.35$ \\

\midrule
\multicolumn{2}{l}{\textit{\textbf{4. OMP, post-processing, \& tracking}}} \\
$G_Z^{\text{gb}},G_X^{\text{gb}}, \Delta_{\text{gb}}$ & $512, 512, 5\lambda$ \\
$G_Z^{\text{refine}},G_X^{\text{refine}}, \Delta_{\text{ref}}$ & $40, 40, \lambda$ \\
$G_Z^{\text{neighbor}},G_X^{\text{neighbor}}, \Delta_{\text{neighbor}}$ & $20, 20, 0.5\lambda$ \\
Tracker thresholds $\epsilon_{\mathrm{th}}$ and $P_{\mathrm{fa}}$ & $0.15, 0.05$ \\
\bottomrule
\end{tabular}
\end{table}

\begin{figure*}[t]
\centering
\subfigure[Proposed, $N_{\text{RF}} \le 512$]{%
    \includegraphics[width=0.13\linewidth]{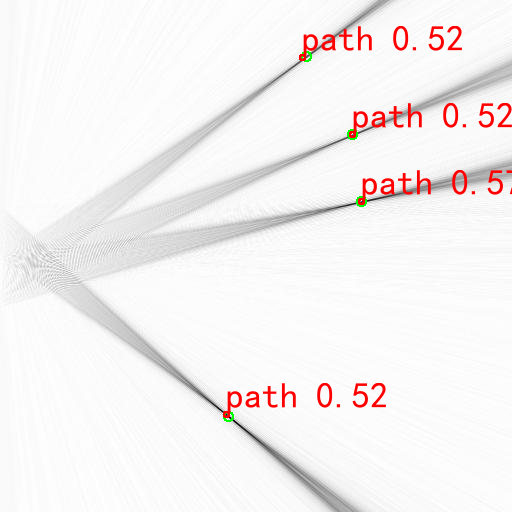}%
    \label{fig:detect0}%
}\hspace{2pt}
\subfigure[Baseline 2, $N_{\text{RF}} = 512$]{%
    \includegraphics[width=0.13\linewidth]{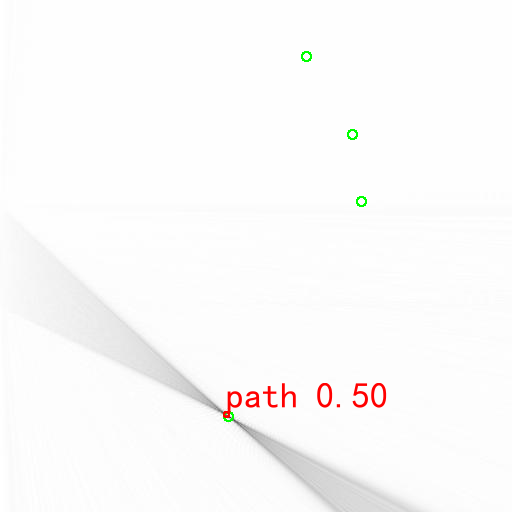}%
    \label{fig:detect1}%
}\hspace{2pt}
\subfigure[Baseline 2, $N_{\text{RF}} = 256$]{%
    \includegraphics[width=0.13\linewidth]{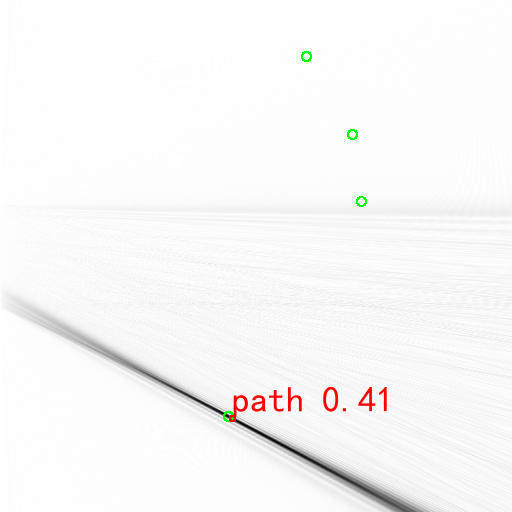}%
    \label{fig:detect3}%
}\hspace{2pt}
\subfigure[Baseline 2, $N_{\text{RF}} = 128$]{%
    \includegraphics[width=0.13\linewidth]{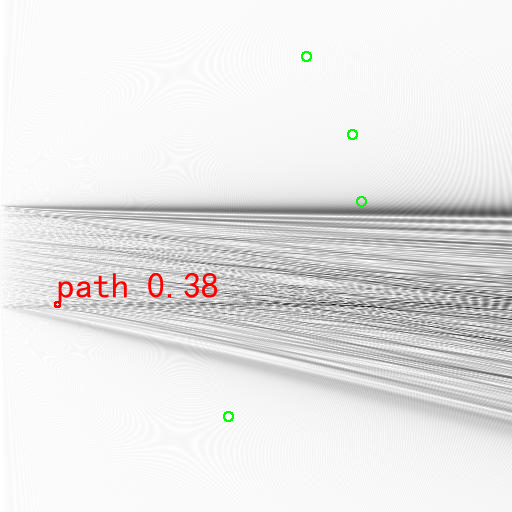}%
    \label{fig:detect5}%
}\hspace{2pt}
\subfigure[Baseline 3, $N_{\text{RF}} = 512$]{%
    \includegraphics[width=0.13\linewidth]{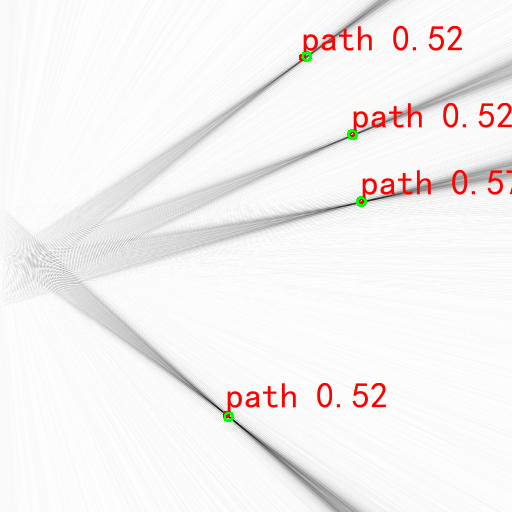}%
    \label{fig:detect2}%
}\hspace{2pt}
\subfigure[Baseline 3, $N_{\text{RF}} = 256$]{%
    \includegraphics[width=0.13\linewidth]{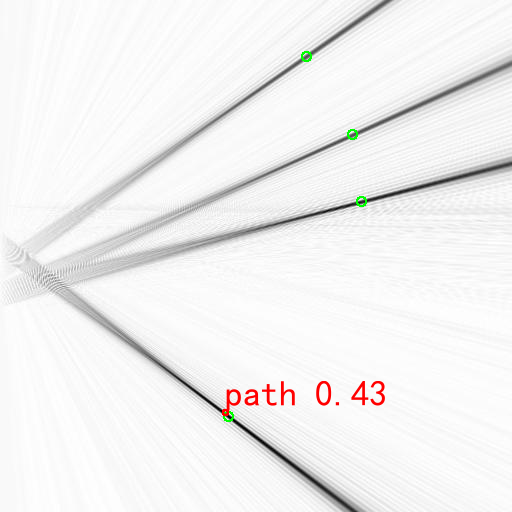}%
    \label{fig:detect4}%
}\hspace{2pt}
\subfigure[Baseline 3, $N_{\text{RF}} = 128$]{%
    \includegraphics[width=0.13\linewidth]{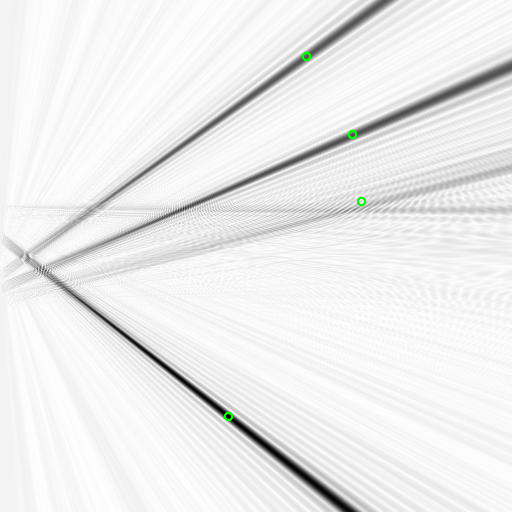}%
    \label{fig:detect6}%
}

\caption{{\bl Detection performance of the improved CenterNet under different HBF schemes (SNR=9dB). The numbers labeled on the detected paths denote the predicted confidence scores. Red and green circles represent estimated and GT positions, respectively.}}
\label{Results_detection_imgs_vs_HBF}
\end{figure*}

\begin{table*}[t]
\caption{Comparison of propagation path detection accuracy under different HBF schemes with different RF chain budgets.}
\label{tab:rf_chain_comparison}
\centering
\renewcommand{\arraystretch}{1.2}
\setlength{\tabcolsep}{3pt}
\begin{tabular}{l ccc c ccc c ccc}
\toprule
\multirow{2}{*}{\textbf{HBF Scheme}} & \multicolumn{3}{c}{\textbf{$N_{\text{RF}} = 512$}} & & \multicolumn{3}{c}{\textbf{$N_{\text{RF}} = 256$}} & & \multicolumn{3}{c}{\textbf{$N_{\text{RF}} = 128$}} \\
\cmidrule{2-4} \cmidrule{6-8} \cmidrule{10-12}
& Recall & Precision & mAP & & Recall & Precision & mAP & & Recall & Precision & mAP \\
\midrule
\multicolumn{12}{l}{\textit{Benchmark: Full Digital ($N_{\text{RF}}=1024$) \hfill Recall: 0.9863 \quad Precision: 0.9742 \quad mAP: 0.9772}} \\
\midrule

Baseline 2 & 0.4912 & 0.4504 & 0.4701 & & 0.2412 & 0.2137 & 0.2212 & & 0.1214 & 0.1012 & 0.1126 \\

Baseline 3 & 0.9831 & 0.9742 & 0.9713 & & 0.8124 & 0.7948 & 0.8037 & & 0.7122 & 0.6851 & 0.6925 \\

\midrule
\textbf{Proposed TM-PC HBF} & \multicolumn{11}{c}{\textbf{Recall: 0.9831 \qquad Precision: 0.9742 \qquad mAP: 0.9713 \quad (Consistent across all $N_{\text{RF}}$)}} \\
\bottomrule
\end{tabular}
\end{table*}

\subsection{Experimental Setups and Baseline Algorithms}
The experimental settings are summarized in Table~\ref{tab:simulation_params}. The training is conducted on a workstation equipped with two NVIDIA GeForce RTX 4090 GPUs, while the testing is performed on a server with an AMD EPYC 7542 32-Core Processor CPU. {\bl To benchmark the proposed signal image-based framework against traditional signal-level processing algorithms, we consider several leading baselines for both the estimation and tracking stages. For the estimation stage, we include typical compressed sensing-based channel estimation methods, i.e., classical OMP~\cite{Lee_TCOM_ChannelEstimation_2016} (\textit{OMP-EST}) and the high-precision NOMP method using near-field Cartesian-domain (\textit{NOMP-EST})~\cite{nomp}, the advanced off-grid polar-domain simultaneous iterative gridless weighted (\textit{P-SIGW-EST}) method~\cite{SIGW}, and the proposed \textit{improved CenterNet-EST} without OMP refinement. For the tracking stage, we consider two representative tracking paradigms, including \emph{path-parameter-based tracking}, represented by NOMP-tracking (\textit{NOMP-TRK})~\cite{extendedNOMPbasedTracking}, and \emph{Bayesian recursive tracking} methods, represented by Extended Kalman-tracking (\textit{EKF-TRK}) and Particle Filter-tracking (\textit{PF-TRK})~\cite{Guerra_TSP_NFTracking_2021}. The proposed method is compared against these baselines in terms of both trajectory and channel tracking accuracy.}

\subsection{Comparison of HBF schemes under Hardware Constraints}

{\bl We first evaluate the impact of HBF designs under varying RF-chain budgets $N_{\text{RF}}$ by examining the quality of the generated signal images and the subsequent path detection accuracy of the improved CenterNet, utilizing the full-digital architecture ($N_{\text{RF}}=1024$) as the performance upper bound. As visually presented in Fig.~\ref{Results_detection_imgs_vs_HBF} and quantitatively supported by the detection evaluation in Table~\ref{tab:rf_chain_comparison} (measured by precision for accuracy, recall for detection completeness, and overall mAP), a decreasing $N_{\text{RF}}$ severely degrades the baseline schemes. Specifically, Baseline~2 exhibits a pronounced performance drop due to expanded angular blind zones that cause paths to be entirely missed. Meanwhile, Baseline~3 suffers from aperture loss and reduced coherent gain. This blurs the convergence zone, yields more false alarms (see Figs.~\ref{fig:detect2}--\ref{fig:detect3}), and degrades its mean Average Precision (mAP) to $0.6925$. In contrast, the proposed TM-PC HBF demonstrates remarkable robustness. By effectively utilizing time-multiplexed combining, it precisely preserves clear ``X-shaped'' propagation-path signatures and yields theoretically equivalent synthesized observations under the intra-slot quasi-static assumption across $N_{\text{RF}} \in \{128,256,512\}$ (Fig.~\ref{fig:detect1}). Specifically, the proposed method consistently achieves a high Recall of 0.9831 and a Precision of 0.9742, indicating exceptionally low missed detection and false alarm rates. Furthermore, with a mean Average Precision (mAP) of 0.9713, it closely approaches the full-digital benchmark. Consequently, the proposed scheme with significantly fewer RF chains successfully matches the performance of the Baseline~3 equipped with $N/2$ RF chains. These quantitative results demonstrate its capability to effectively mitigate RF hardware constraints while ensuring high-fidelity path detection.}

\begin{figure}[t]
\centering
\subfigure[SNR=18 dB]{%
    \includegraphics[width=0.3\linewidth]{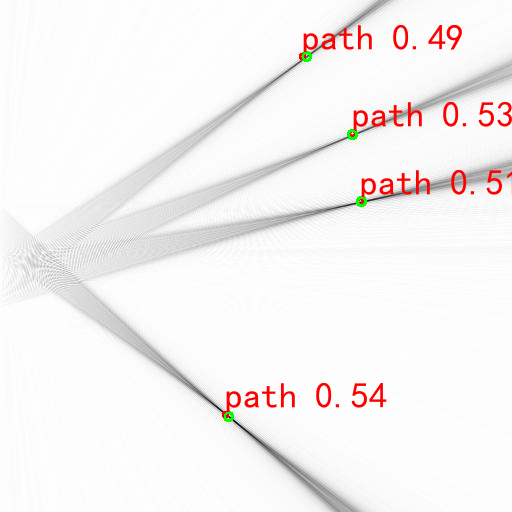}%
    \label{fig:detect7}%
}\hspace{0pt}
\subfigure[SNR=9 dB]{%
    \includegraphics[width=0.3\linewidth]{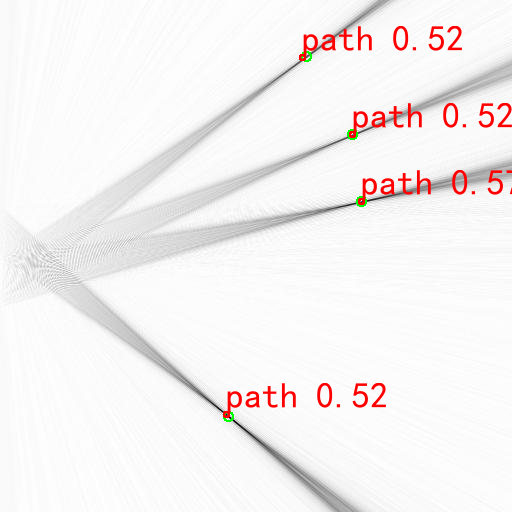}%
    \label{fig:detect8}%
}\hspace{0pt}
\subfigure[SNR=-3 dB]{%
    \includegraphics[width=0.3\linewidth]{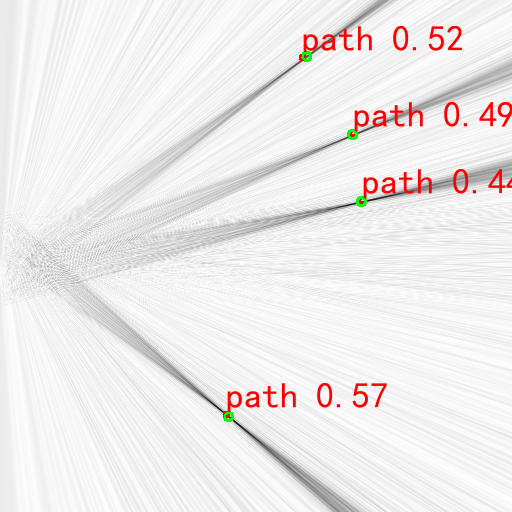}%
    \label{fig:detect9}%
}

\caption{{\bl Detection performance under different SNRs.}}
\label{Results_detection_imgs_vs_SNR}
\end{figure}

{\bl
Furthermore, we evaluate the robustness of the proposed HBF scheme under various SNRs.  As shown in Fig.~\ref{Results_detection_imgs_vs_SNR}, the proposed method maintains stable keypoint detection performance as the SNR drops from $18$~dB to $-3$~dB, experiencing some marginal confidence score degradations. This reliability can be attributed to the high-fidelity signal images, which effectively preserve distinct spatial path characteristics even under challenging noise conditions. }

\subsection{Performance of Initial Time Slot Estimation}

The initial estimation performance of the proposed framework is evaluated from two perspectives: position estimation of propagation paths and channel estimation. The position estimation error is quantified by the root mean square error (RMSE), defined as $\mathrm{RMSE} = \sqrt{\mathbb{E}\{\|\hat{\mathbf{p}} - \mathbf{p}\|^2\}}$ (m), where $\mathbf{p}$ and $\hat{\mathbf{p}}$ represent GT and estimated path positions, respectively.
Meanwhile, the channel estimation accuracy is measured by the normalized mean square error (NMSE), defined as $\mathrm{NMSE}= 10\log_{10}\mathbb{E} \{\|\hat{\mathbf{h}}-\mathbf{h}\|^2 \, /\|\mathbf{h}\|^{2} \}$ (dB), where $\mathbf{h}$ and $\hat{\mathbf{h}}$ denote the GT and estimated channels. {\bl Moreover, the Cram\'er--Rao lower bound (CRLB) is widely adopted as a fundamental theoretical benchmark for evaluating channel estimation algorithms~\cite{Kay_Estimation_1993}. To provide a corresponding theoretical reference for both localization and channel estimation performance, we further derive the CRLBs for path-position parameters and channels under the proposed HBF scheme, as detailed in Appendix~\ref{appendix:crlb}.}

\begin{figure}[t]
    \centering
    \subfigure[Position estimation performance.]{
        \includegraphics[width=0.7\linewidth]{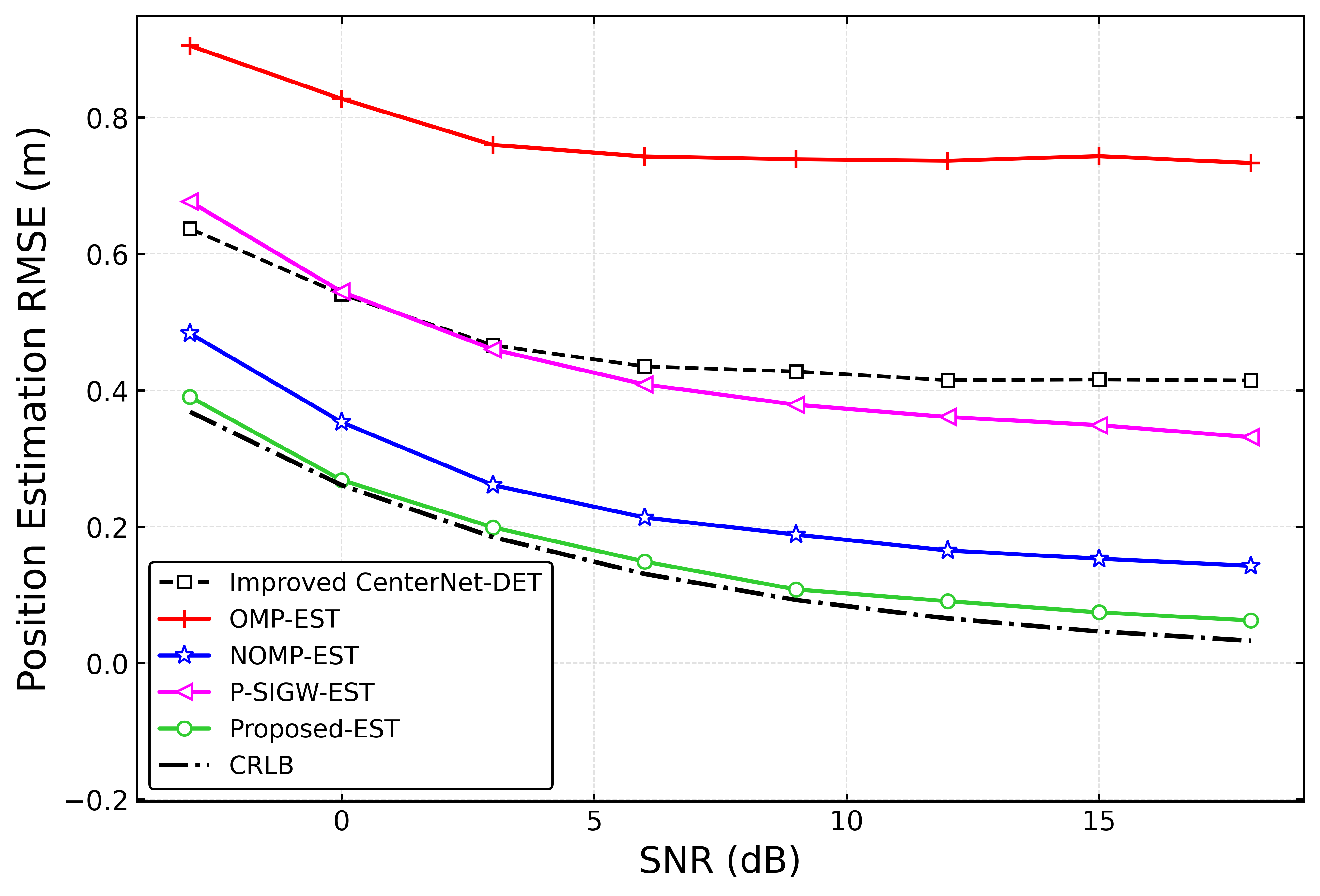}
        \label{fig:EST_POS}
    }
    
    \subfigure[Channel estimation performance.]{
        \includegraphics[width=0.7\linewidth]{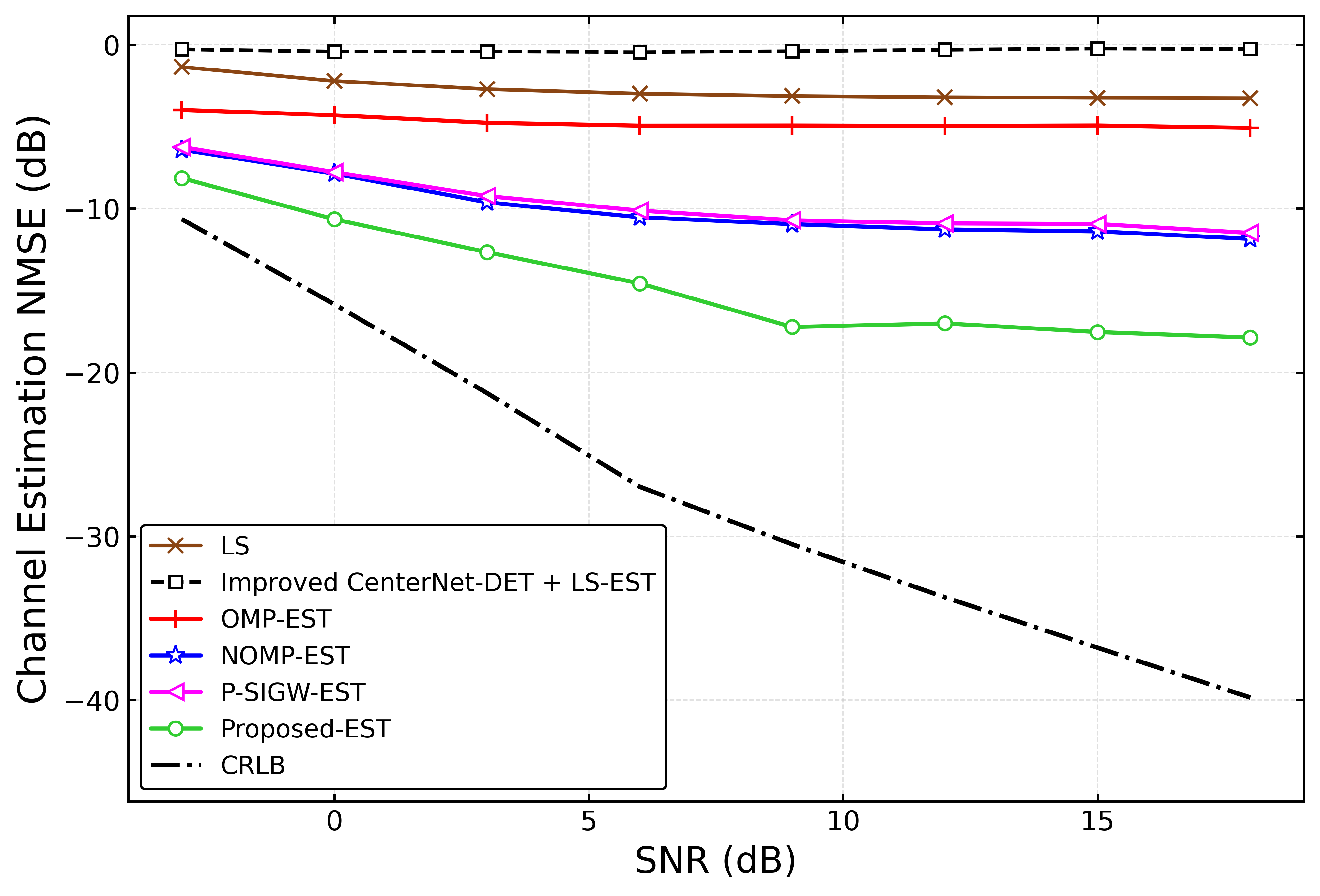}
        \label{fig:EST_CHANNEL}
    }

    \caption{{\bl Estimation performance across varying SNRs under the proposed TM-PC HBF scheme at the initial time slot ($N_{\text{RF}}=128$).}}
    \label{fig:estimation_performance}
\end{figure}


\begin{figure*}[t]
\centering
\subfigure[Trajectories]{%
    \includegraphics[width=0.138\linewidth]{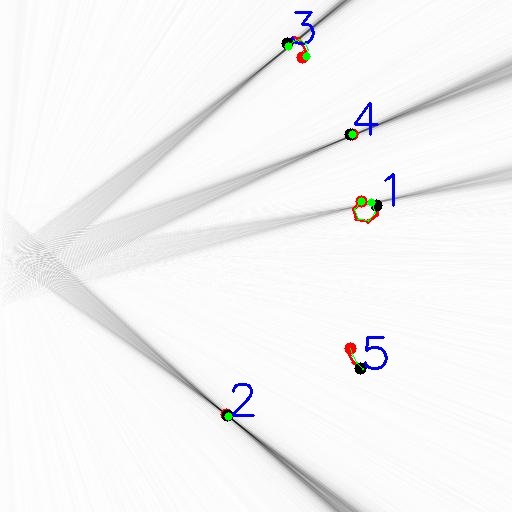}%
    \label{fig:traj_all}%
}\hspace{2pt}
\subfigure[Time slot 1]{%
    \includegraphics[width=0.138\linewidth]{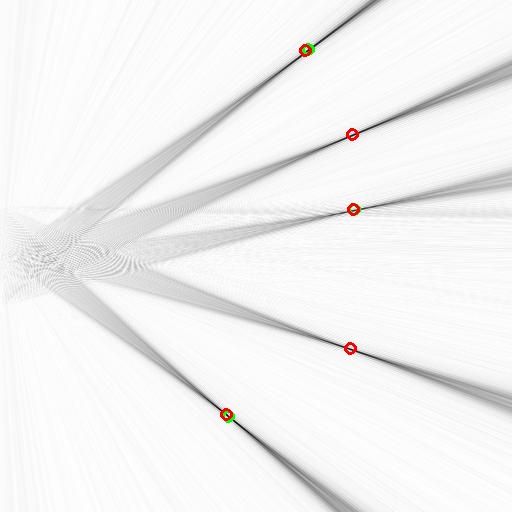}%
    \label{fig:pos_t1}%
}\hspace{2pt}%
\subfigure[Time slot 2]{%
    \includegraphics[width=0.138\linewidth]{Figs/Results/r_trajectory/frame_1.jpg}%
    \label{fig:pos_t2}%
}\hspace{2pt}%
\subfigure[Time slot 3]{%
    \includegraphics[width=0.138\linewidth]{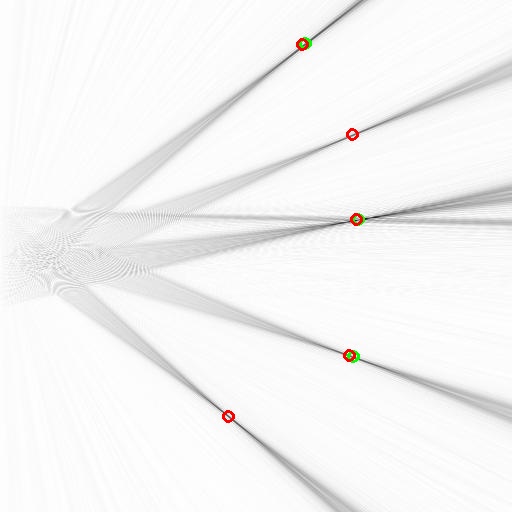}%
    \label{fig:pos_t3}%
}\hspace{2pt}%
\subfigure[Time slot 4]{%
    \includegraphics[width=0.138\linewidth]{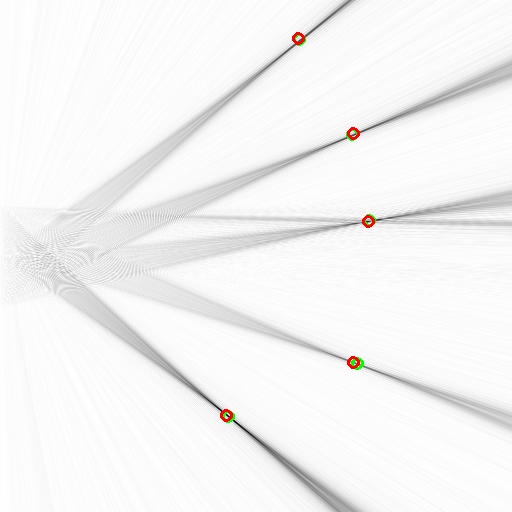}%
    \label{fig:pos_t4}%
}\hspace{2pt}%
\subfigure[Time slot 5]{%
    \includegraphics[width=0.138\linewidth]{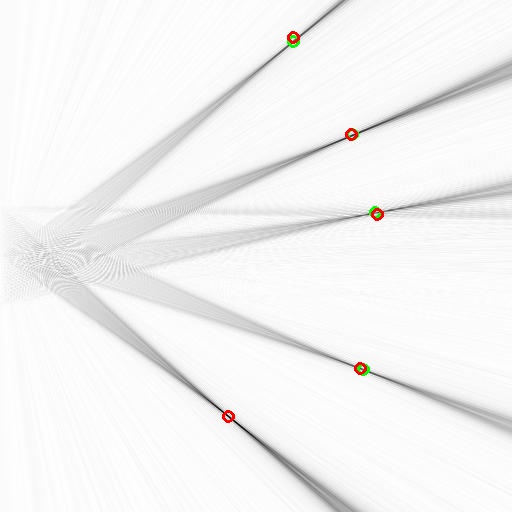}%
    \label{fig:pos_t5}%
}\hspace{2pt}%
\subfigure[Time slot 6]{%
    \includegraphics[width=0.138\linewidth]{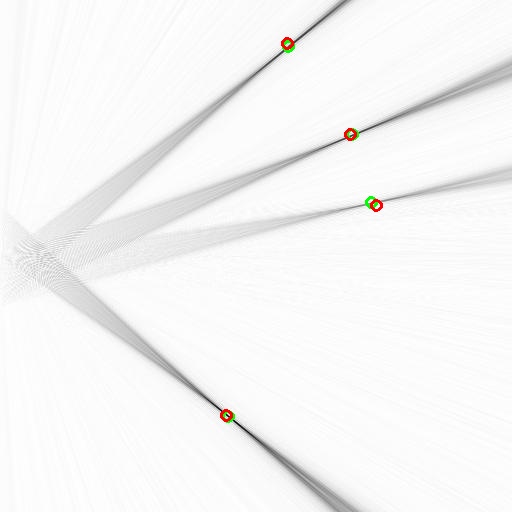}%
    \label{fig:pos_t6}%
}

\caption{{\bl Trajectory tracking results visualization (SNR=9 dB). (a) The overall estimated trajectories (red lines) and corresponding GTs (green lines), labeled with their respective path IDs (blue numbers). The start and end points of trajectories are marked in black and red, respectively. (b)–(g) Snapshot localization results across the observed time slots.}}
\label{Results_monitoring_time}
\end{figure*}

\begin{figure*}[t]
\centering
\subfigure[SNR = -3 dB]{
    \includegraphics[width=0.15\linewidth]{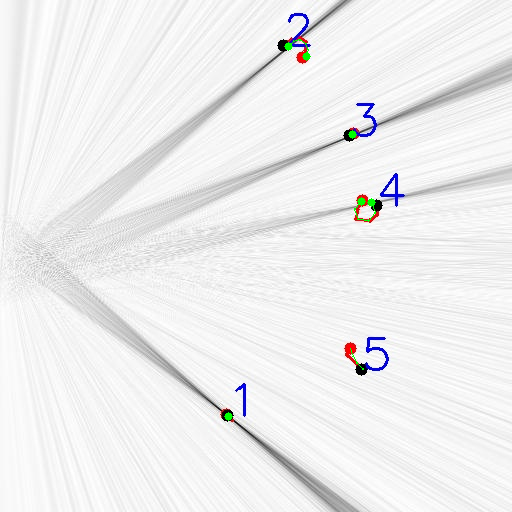}
    \label{fig:Trajectory_-3dB}
}
\subfigure[SNR = 3 dB]{
    \includegraphics[width=0.15\linewidth]{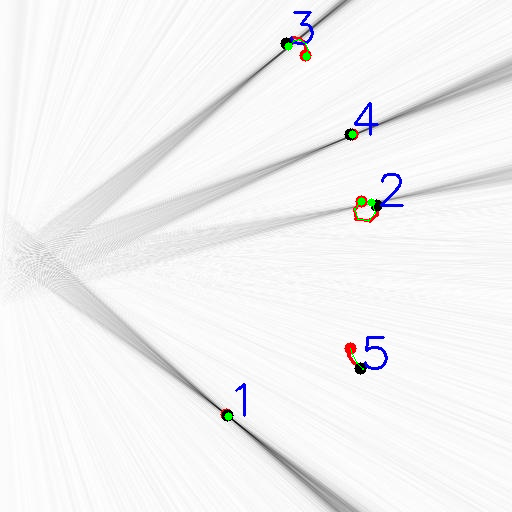}
    \label{fig:Trajectory_3dB}
}
\subfigure[SNR = 9 dB]{
    \includegraphics[width=0.15\linewidth]{Figs/Results/r_trajectory/Trajectory_9_2.png}
    \label{fig:Trajectory_9dB}
}
\subfigure[SNR = 12 dB]{
    \includegraphics[width=0.15\linewidth]{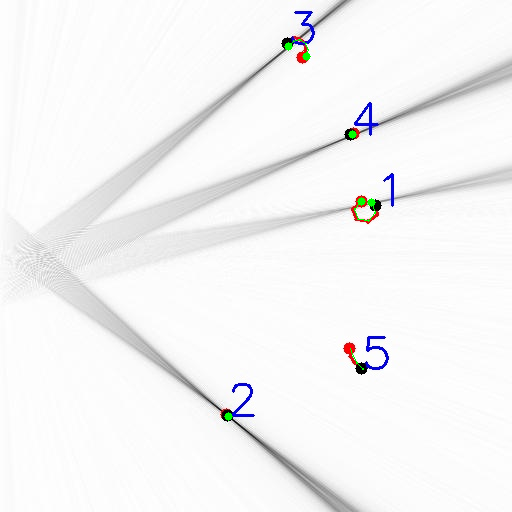}
    \label{fig:Trajectory_12dB}
}
\subfigure[SNR = 18 dB]{
    \includegraphics[width=0.15\linewidth]{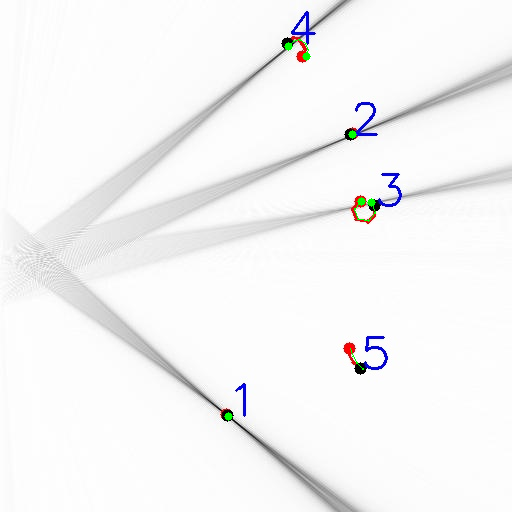}
    \label{fig:Trajectory_18dB}
}
\caption{Trajectory tracking results under different SNRs.}
\label{Results7_trajectories}
\end{figure*}

Figs.~\ref{fig:estimation_performance} shows the position RMSE and channel NMSE versus SNR, respectively. As shown in Fig.~\ref{fig:EST_POS}, the improved CenterNet alone achieves a lower position estimation error than OMP-EST and provides comparable performance to P-SIGW-EST. This performance gain mainly comes from the offset prediction of the improved Centernet, which effectively compensates for the quantization error introduced by discrete dictionary sampling of OMP. {\bl Building on this high-quality coarse localization, the proposed algorithm further improves the position estimation accuracy, consistently outperforming NOMP-EST and approaching the CRLB over a wide SNR range, which demonstrates the high localization accuracy of the proposed framework.}

For channel estimation, Fig.~\ref{fig:EST_CHANNEL} reveals a notable performance inversion that although the standalone improved CenterNet excels in spatial localization, the detection with LS gain estimator yields an inferior NMSE compared to OMP-EST. This highlights that purely geometric detection is insufficient to resolve fine-grained complex phases and weak multipath components. However, leveraging the highly accurate spatial anchors already provided by CenterNet, the proposed OMP refiner only needs to operate within a narrowly confined, small-scale codebook to rapidly acquire precise channel parameters and reconstruct the channel. By mitigating the phase bottleneck with such a lightweight mechanism, the complete framework ultimately outperforms both NOMP-EST and P-SIGW-EST by approximately $2$--$10$~dB across all SNRs, validating the necessity and efficacy of the detection-and-refinement architecture.

\subsection{Performance of Overall Tracking}

{\bl Fig.~\ref{Results_monitoring_time} demonstrates that the proposed framework can reliably support trajectory-level environment monitoring at $\mathrm{SNR}=9$~dB. In particular, IDs~2 and~4 are correctly identified as static scatterers, while the dynamic trajectories of IDs~1, 3, and~5 remain closely aligned with the GT. Moreover, the disappearance of ID~5 is clearly captured, verifying the capability of the proposed method to detect trajectory birth--death events.}
As shown in Fig.~\ref{Results7_trajectories}, the estimated trajectories generally agree well with the GTs across all SNRs, indicating that the proposed method can reliably recover the main paths of both the user and scatterers even at low SNR. As the SNR increases, the estimated paths become more tightly aligned with the GTs, further demonstrating improved localization and tracking accuracy.

\begin{figure}[t]
    \centering
    \subfigure[Trajectory tracking performance.]{
        \includegraphics[width=0.65\linewidth]{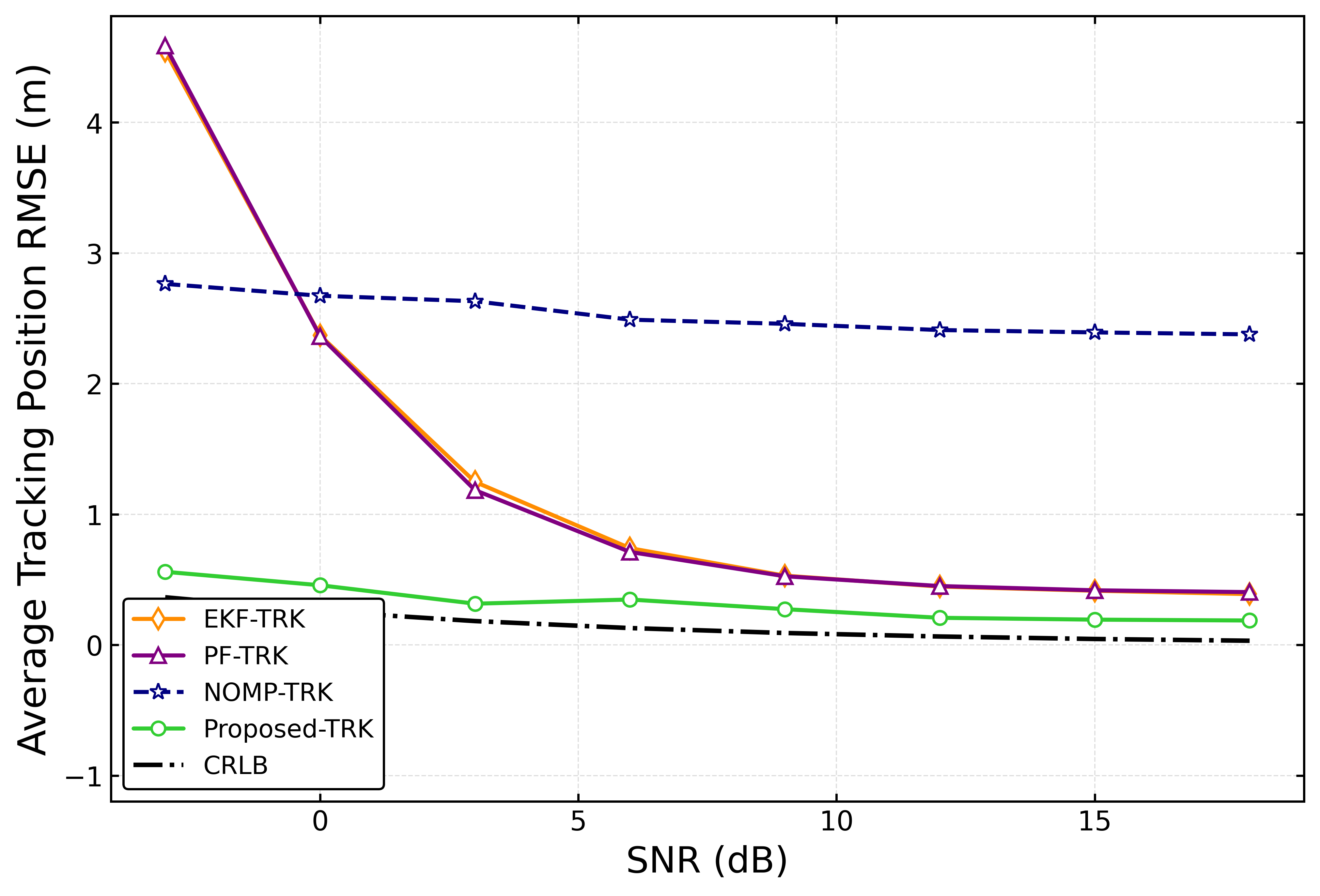}
        \label{fig:TRK_POS_snr}
    }
    
    \subfigure[Channel tracking performance.]{
        \includegraphics[width=0.65\linewidth]{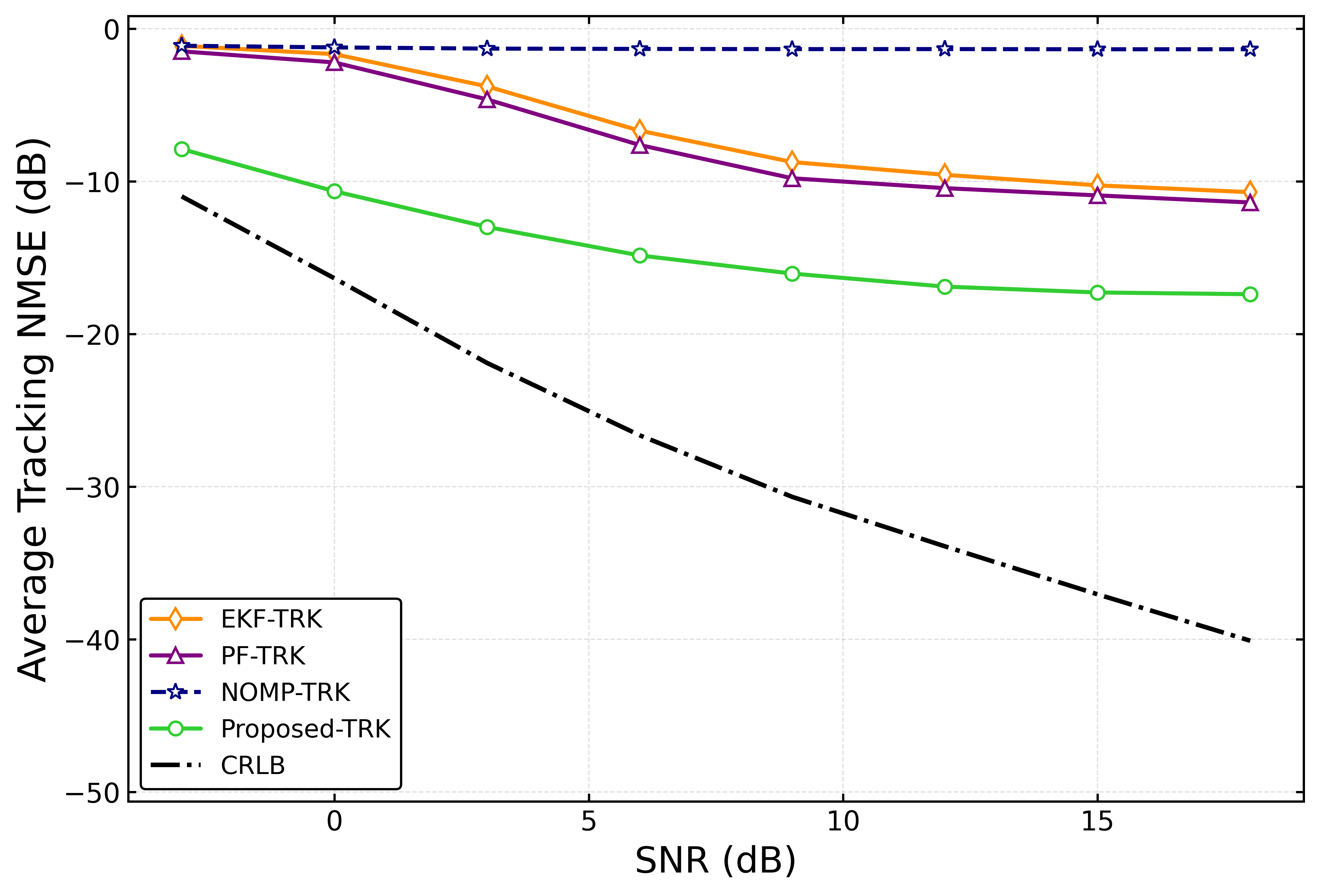}
        \label{fig:TRK_CHANNEL_snr}
    }

    \caption{{\bl Tracking performance versus SNR under the proposed TM-PC HBF scheme.}}
    \label{fig:tracking_performance_snr}
\end{figure}

\begin{figure}[t]
    \centering
    \subfigure[Trajectory tracking performance.]{
        \includegraphics[width=0.65\linewidth]{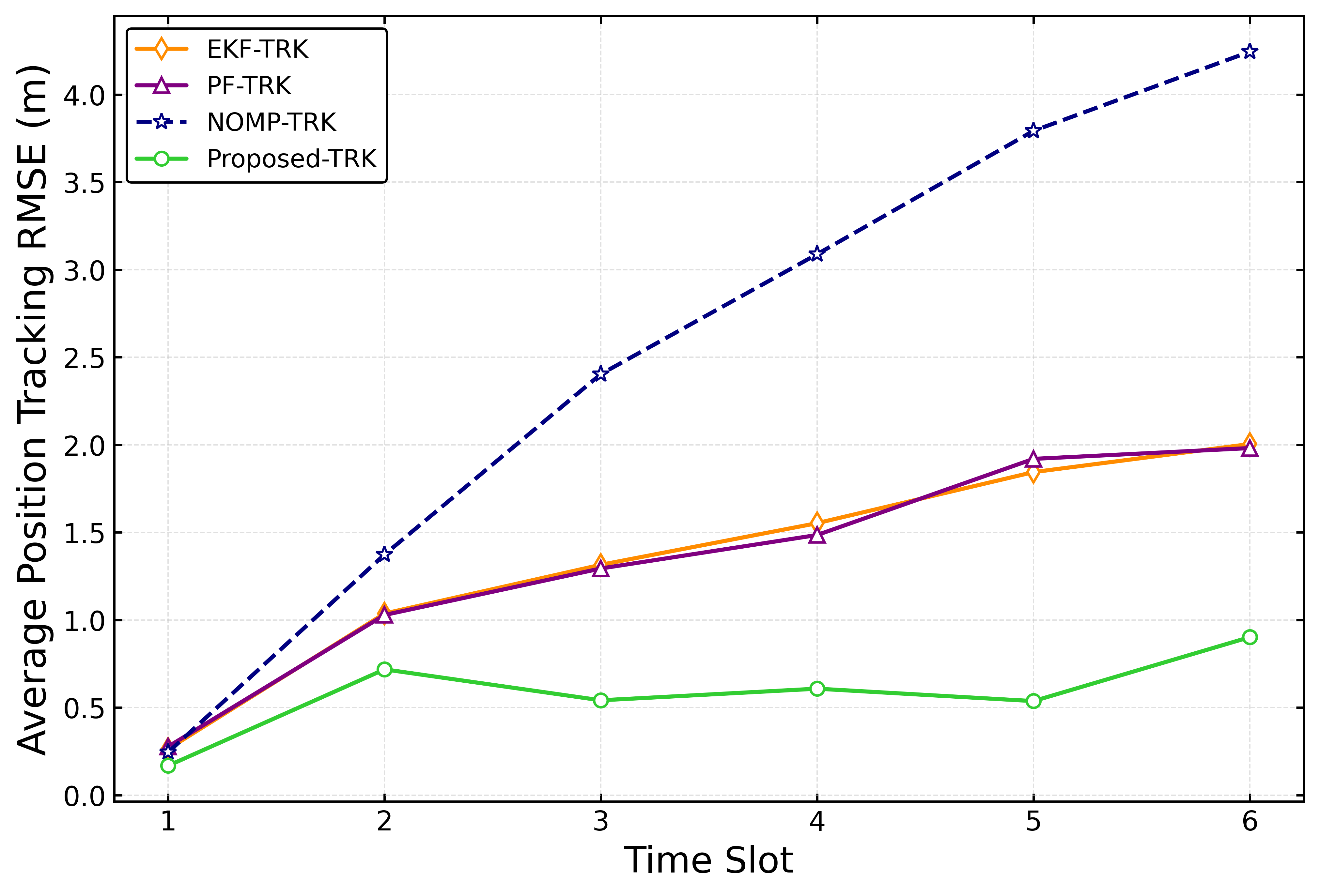}
        \label{fig:TRK_POS}
    }
    
    \subfigure[Channel tracking performance.]{
        \includegraphics[width=0.65\linewidth]{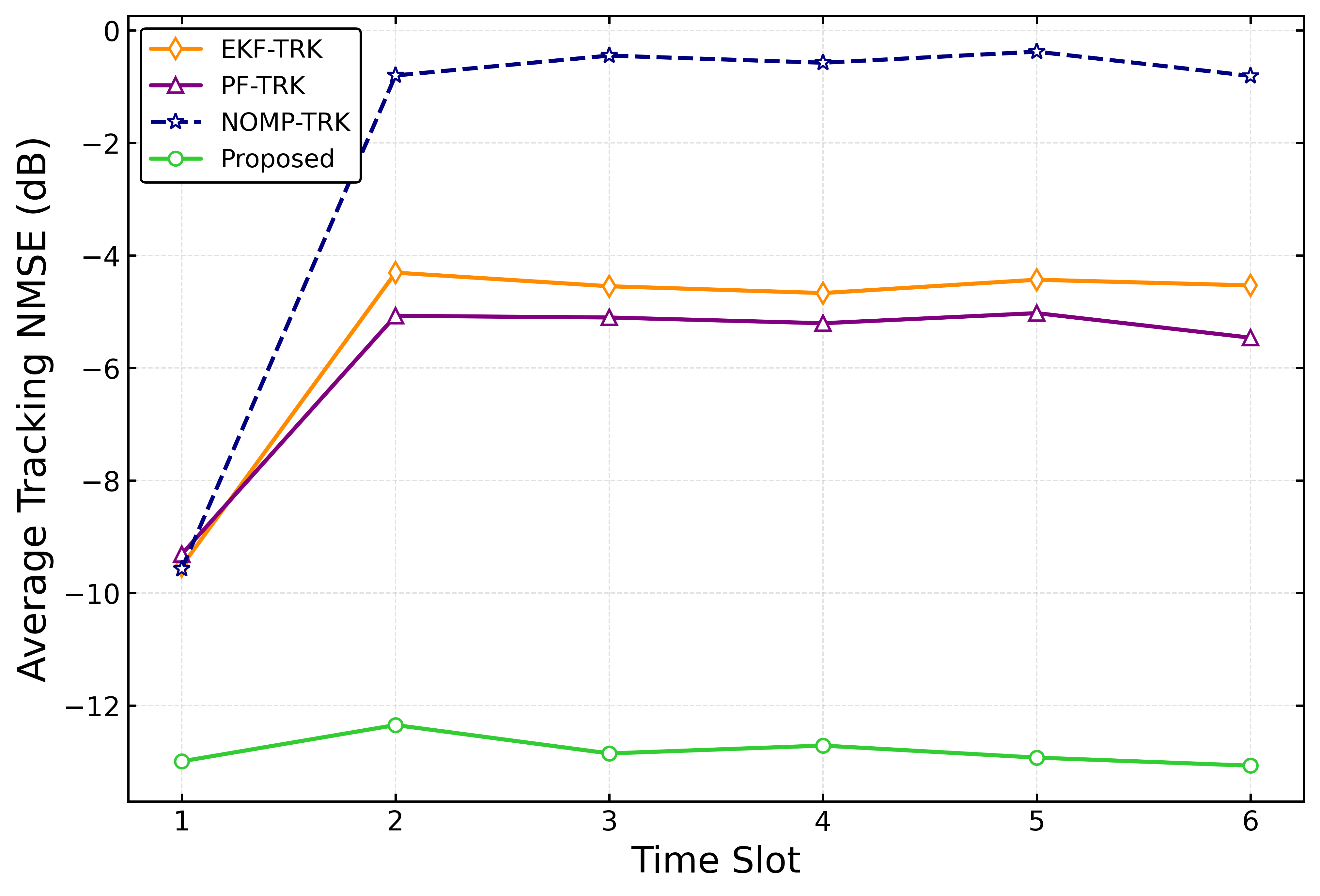}
        \label{fig:TRK_CHANNEL}
    }

    \caption{{\bl Tracking performance over consecutive time slots under the proposed TM-PC HBF scheme.}}
    \label{fig:tracking_performance}
\end{figure}

\begin{figure}[t]
    \centering
    \subfigure[Sample 1.]{
        \includegraphics[width=0.46\linewidth]{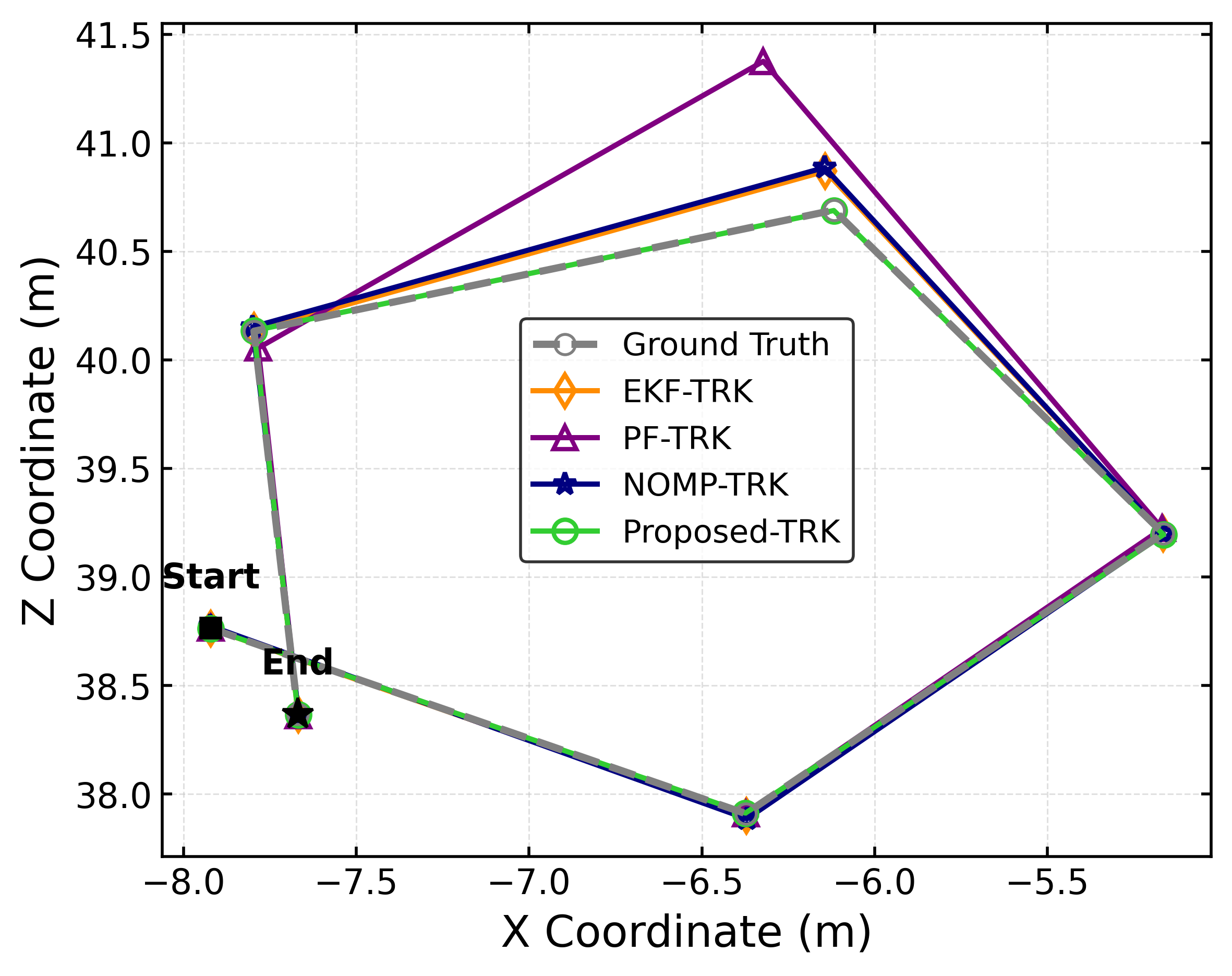}
        \label{fig:track_vis1}
    }
    \subfigure[Sample 2.]{
        \includegraphics[width=0.46\linewidth]{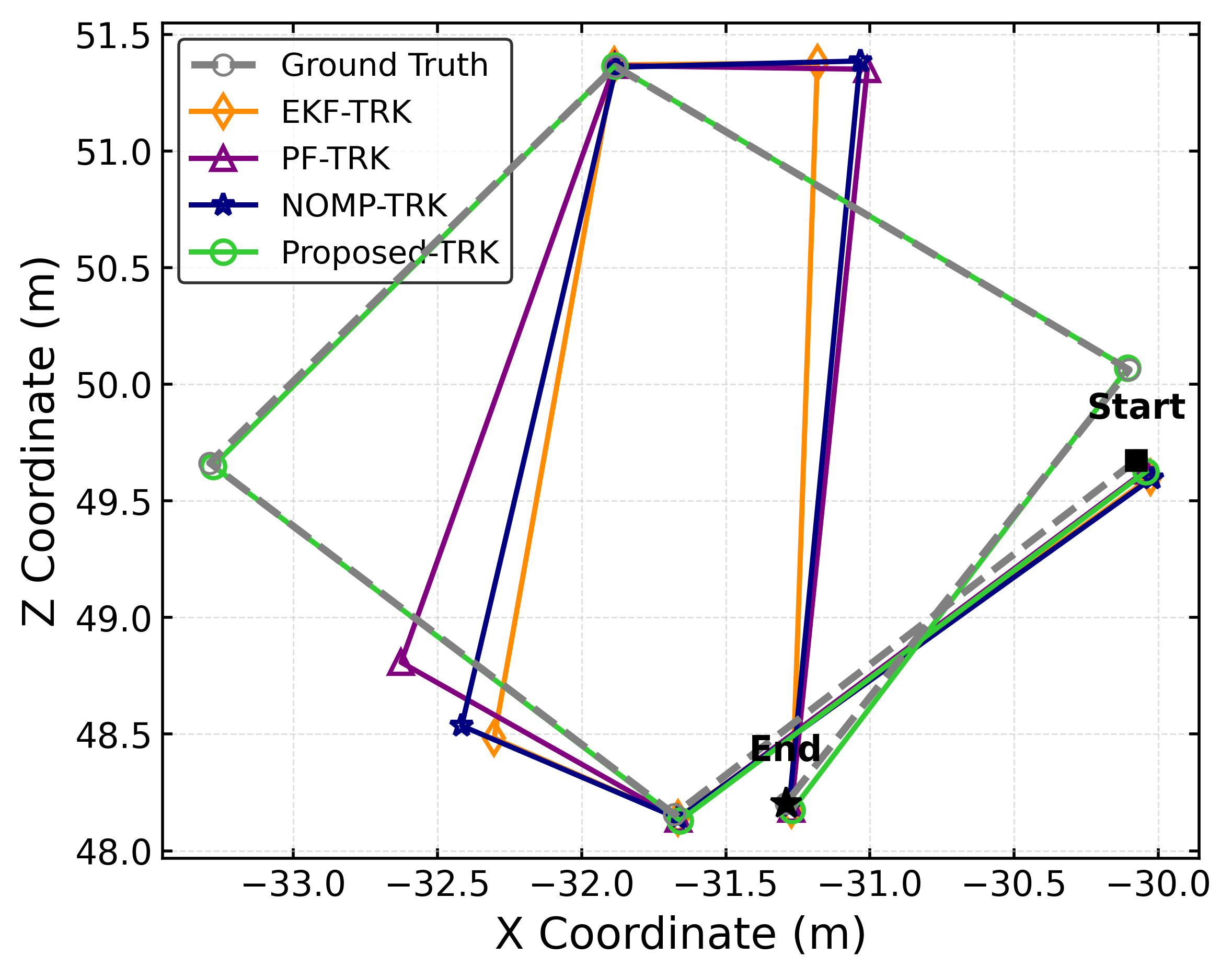}
        \label{fig:nmse_dist}
    }
    \caption{{\bl Tracked trajectories comparison of different algorithms (SNR=9 dB).}}
    \label{fig:track_vis2}
\end{figure}

{\bl As shown in Fig.~\ref{fig:TRK_POS_snr}, the proposed framework consistently achieves the lowest position tracking RMSE across the entire SNR range and remains relatively stable as the SNR varies, demonstrating strong robustness to noise. In contrast, EKF-TRK and PF-TRK suffer from severe performance degradation in the low-SNR regime, although their errors gradually decrease as the SNR increases, while NOMP-TRK remains significantly less accurate over all SNRs. A similar trend can also be observed in Fig.~\ref{fig:TRK_CHANNEL_snr} for channel tracking.
Fig.~\ref{fig:tracking_performance} illustrates the tracking accuracy for both position and channel over $T=6$ consecutive observed time slots, with the results averaged across all considered SNR levels. As observed in both Fig.~\ref{fig:TRK_POS} and Fig.~\ref{fig:TRK_CHANNEL}, the proposed framework consistently outperforms the baseline algorithms, i.e., EKF-TRK, PF-TRK, and NOMP-TRK. Furthermore, while the tracking errors of the baseline methods tend to increase over time, the proposed method demonstrates remarkable stability. 
To provide an intuitive comparison, Fig.~\ref{fig:track_vis2} shows two representative trajectory examples at $\mathrm{SNR}=9$~dB. It can be observed that Proposed-TRK remains closest to the GT throughout the whole trajectory, whereas EKF-TRK, PF-TRK, and NOMP-TRK exhibit progressively larger deviations as time evolves, indicating stronger error accumulation.}

\begin{figure}[t]
    \centering
    \subfigure[Trajectory tracking performance.]{
        \includegraphics[width=0.65\linewidth]{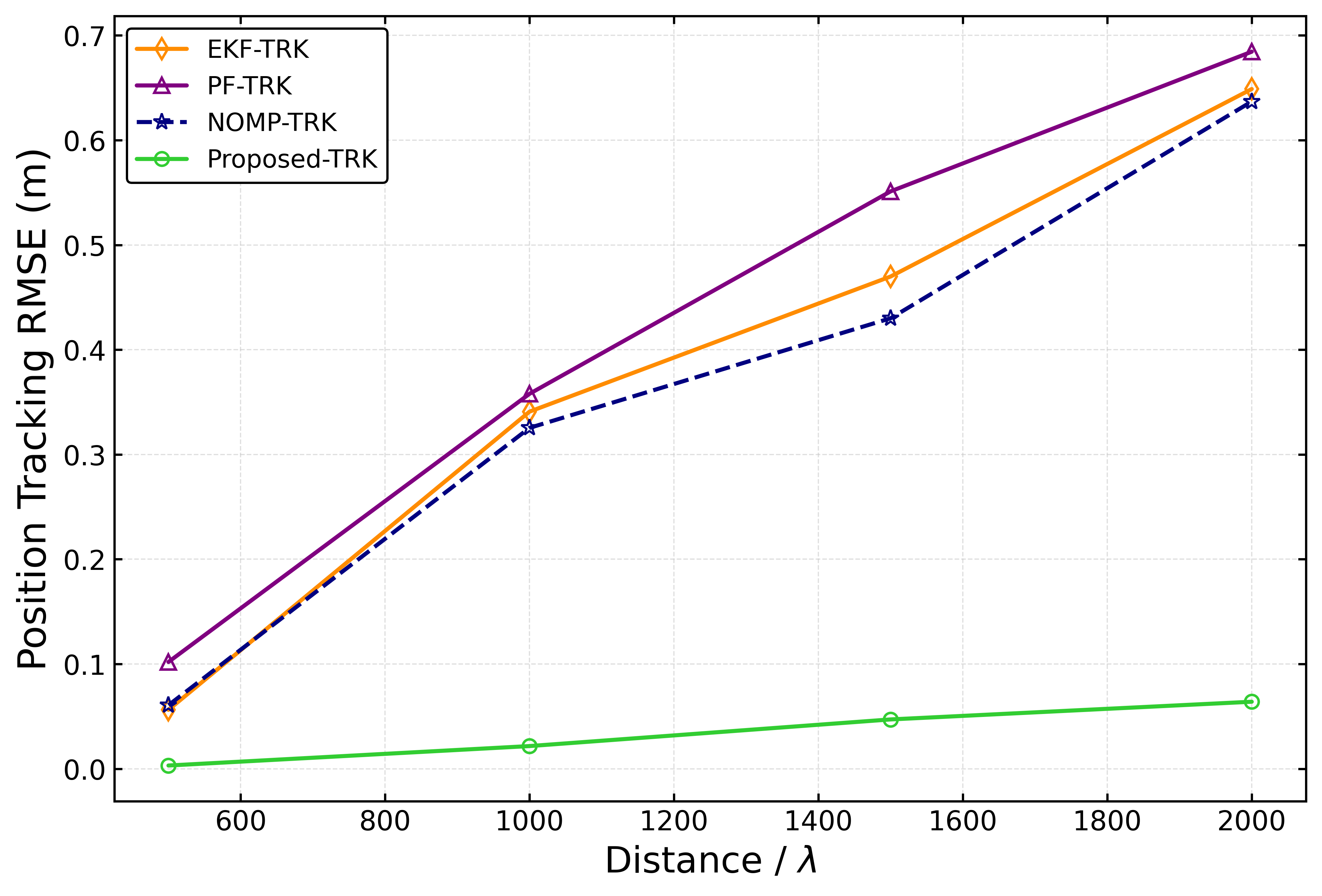}
        \label{fig:dis_pos_TRK}
    }
    
    \subfigure[Channel tracking performance.]{
        \includegraphics[width=0.65\linewidth]{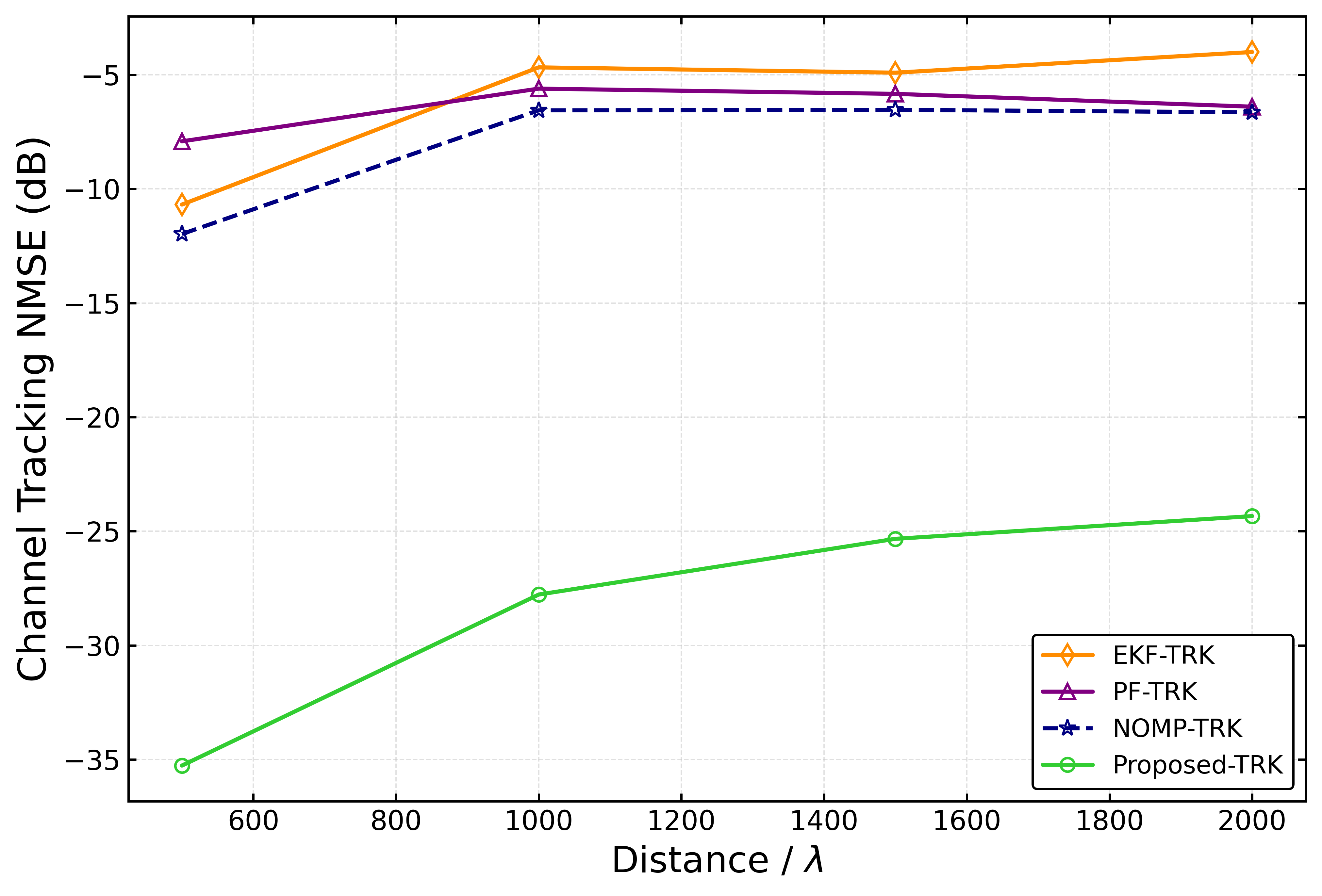}
        \label{fig:dis_channel_TRK}
    }

    \caption{{\bl Tracking performance of LoS paths across varying distances between the UE and BS.}}
    \label{fig:performance_distance}
\end{figure}

{\bl Finally, in near-field regions, propagation distance is also a crucial factor influencing tracking accuracy. Therefore, we evaluate the average tracking performance (averaged across all SNRs and time slots) versus the UE-BS distance, utilizing channels and signals generated in the scenario of a mobile user with a Line-of-Sight (LoS) path as an example. Utilizing a comprehensive set of 4,800 data samples, we investigate the overall tracking performance (averaged across all SNRs and time slots) versus the UE-BS distance. As depicted in Fig.~\ref{fig:dis_pos_TRK} and Fig.~\ref{fig:dis_channel_TRK}, the tracking errors for both position and channel estimation grows for all algorithms as the distance extends from $500\lambda$ to $2000\lambda$. However, while the position RMSE of the baselines suffers from a severe deterioration, the Proposed-TRK exhibits a significantly smaller degradation margin. Despite the increasing distance, it demonstrates exceptional robustness by bounding the position error strictly below 0.1~m. A consistent channel tracking trend is observed in Fig.~\ref{fig:dis_channel_TRK} that although the NMSE of the proposed method gradually increases at larger distances, its error growth rate is substantially slower than that of the baselines, ensuring the NMSE remains well below $-25$~dB. Overall, the proposed framework effectively curbs the severe distance-induced performance degradation inherent in conventional algorithms.}


\begin{table}[t]
\centering
\caption{{\bl Complexity comparison of representative estimation and tracking schemes.}}
\label{tab:complexity_comparison}
\renewcommand{\arraystretch}{1.12}
\setlength{\tabcolsep}{2.5pt}
\scriptsize
\resizebox{\columnwidth}{!}{%
\begin{tabular}{lcc}
\toprule
\textbf{Scheme} & \textbf{Stage} & \textbf{Complexity} \\
\midrule
OMP-EST 
& Est. 
& $\mathcal{O}\!\left(\frac{LN}{2}|\mathcal{G}_{\mathrm{gb}}|\right)$ \\

NOMP-EST/SIGW-EST
& Est. 
& $\mathcal{O}\!\left(\frac{LN}{2}|\mathcal{G}_{\mathrm{gb}}|
+LK_{\mathrm{it}}\left(\frac{N}{2}\right)^2\right)$ \\

NOMP-TRK
& Trk. 
& $\mathcal{O}\!\left(\frac{LN}{2}|\mathcal{G}_{\mathrm{nb}}|
+LK_{\mathrm{it}}^{\mathrm{trk}}\left(\frac{N}{2}\right)^2\right)$ \\

EKF-TRK
& Trk. 
& $\mathcal{O}\!\left(\frac{LN}{2}
\left(|\mathcal{G}_{\mathrm{nb}}|+|\mathcal{G}_{\mathrm{ref}}|\right)\right)+\mathcal{O}(L)$ \\

PF-TRK
& Trk. 
& $\mathcal{O}\!\left(\frac{LN}{2}
\left(|\mathcal{G}_{\mathrm{nb}}|+|\mathcal{G}_{\mathrm{ref}}|\right)\right)+\mathcal{O}(LN_p)$ \\

\rowcolor{gray!15}
Proposed-EST
& Est. 
& $\mathcal{O}\!\left(\frac{N}{2}|\mathcal{G}_{\mathrm{gb}}|\right)
+C_{\mathrm{CNN}}
+\mathcal{O}\!\left(\frac{LN}{2}|\mathcal{G}_{\mathrm{ref}}|\right)$ \\

\rowcolor{gray!15}
Proposed-TRK
& Trk. 
& $\mathcal{O}\!\left(\frac{LN}{2}
\left(|\mathcal{G}_{\mathrm{nb}}|+|\mathcal{G}_{\mathrm{ref}}|\right)\right)$ \\
\bottomrule
\end{tabular}%
}
\end{table}

{\bl
\subsection{Computational Complexity Analysis}
\label{subsec:complexity}

The complexity comparison is summarized in Table~\ref{tab:complexity_comparison}. $K_{\mathrm{it}}$ and $K_{\mathrm{it}}^{\mathrm{trk}}$ denote the numbers of iterative refinement steps in the initial estimation and tracking stages, respectively, $N_p$ denotes the number of particles used in PF-TRK, and $\rho_{\mathrm{re}}$ denotes the re-detection ratio.
For path-iterative initial estimation baselines, i.e., OMP-EST, NOMP-EST, and SIGW-EST, the complexity scales linearly with the path number $L$ because these methods perform path-wise global search over $\mathcal{G}_{\mathrm{gb}}$. Specifically, the complexity of OMP-EST is $\mathcal{O}\big((LN/2)|\mathcal{G}_{\mathrm{gb}}|\big)$, while NOMP-EST and SIGW-EST further require iterative refinement or off-grid updates with complexity of $\mathcal{O}LK_{\mathrm{it}}\left({N}/{2}\right)^2$.

By contrast, the proposed framework follows a \emph{one-shot detection + local refinement + local tracking} strategy. Its initial estimation complexity is
\begin{equation}
\mathcal{O}_{\mathrm{prop,est}}
=
\mathcal{O}\!\left(\frac{N}{2}|\mathcal{G}_{\mathrm{gb}}|\right)
+
C_{\mathrm{CNN}}
+
\mathcal{O}\!\left(\frac{LN}{2}|\mathcal{G}_{\mathrm{ref}}|\right).
\end{equation}
Since the network architecture and input/output resolutions are fixed, $C_{\mathrm{CNN}}$ is independent of $L$ and $T$, and it is invoked only for initialization or re-detection. Compared with path-wise global search, the proposed method performs the global projection only once and restricts path-wise refinement to the small local grid $\mathcal{G}_{\mathrm{ref}}$.

For subsequent slots, when the residual-based validity check is satisfied, the proposed cascaded OMP tracker directly updates each path from the received signal using previous-slot estimates as spatial priors. Therefore, signal-image generation and CenterNet inference are not required in normal tracking slots. The per-slot tracking complexity is
\begin{equation}
\mathcal{O}_{\mathrm{prop,trk}}
=
\mathcal{O}\!\left(
\frac{LN}{2}
\left(
|\mathcal{G}_{\mathrm{nb}}|
+
|\mathcal{G}_{\mathrm{ref}}|
\right)
\right),
\end{equation}
where $|\mathcal{G}_{\mathrm{nb}}|,|\mathcal{G}_{\mathrm{ref}}|\ll|\mathcal{G}_{\mathrm{gb}}|$. If the validity check fails, the signal image is regenerated and the detector is re-activated and the complexity of this re-detection slot is approximately $\mathcal{O}_{\mathrm{prop,est}}$.
The proposed framework is also more efficient than representative tracking baselines. NOMP-TRK also exploits the previous-slot position but still requires Newton-type iterative local refinement for each path, while PF-TRK further incurs particle propagation and resampling overhead. By contrast, the proposed method only performs lightweight two-stage local matching in normal tracking slots, leading to a significantly lower online tracking burden.

Under the adopted settings, the proposed local refinement reduces the effective search burden by $|\mathcal{G}_{\mathrm{gb}}|/|\mathcal{G}_{\mathrm{ref}}|\approx164\times$ in the initial estimation stage, while normal tracking achieves a per-slot search reduction of $|\mathcal{G}_{\mathrm{gb}}|/(|\mathcal{G}_{\mathrm{nb}}|+|\mathcal{G}_{\mathrm{ref}}|)\approx131\times$. With the tested re-detection ratio $\rho_{\mathrm{re}}=3.2\%$ in our experiments, the overall search-related complexity reduction is approximately $14\times$--$25\times$ for $L\in[3,6]$.
}

\section{Conclusion}
\label{sec:Conclusion}

In this paper, we proposed a vision-based framework for efficient joint trajectory and channel tracking in near-field XL-MIMO systems. To overcome the severe latency of conventional path-iterative search-and-refinement methods, we transformed the received signals into high-quality signal images and subsequently performed fast propagation path detection. To this end, we designed a TM-PC HBF architecture to generate high-fidelity signal images under limited RF-chain budgets. Driven by these images, the improved CenterNet first localized the user and surrounding scatterers in a one-shot manner with high accuracy, requiring only a lightweight local refinement to obtain high-precision path parameters and channel estimates. On this basis, a cascaded OMP tracker further exploited temporal correlation for efficient path updating, while residual-based re-detection ensured robustness against tracking failure and path birth-death. Simulation results demonstrated that the proposed framework achieved superior trajectory and channel tracking accuracy with substantially reduced latency, making it a promising solution for low-complexity near-field XL-MIMO sensing and communications.

{\bl
\appendices

\section{Cram\'{e}r-Rao Lower Bounds}
\label{appendix:crlb}

This appendix derives the CRLBs for multi-path position and channel reconstruction under the proposed hybrid beamforming architecture. For brevity, the time-slot index $t$ is omitted.

\subsection{CRLB for Multi-Path Position Estimation}

We use $\boldsymbol{\eta}=[\boldsymbol{\eta}_1^{\top},\dots,\boldsymbol{\eta}_L^{\top}]^{\top}\in\mathbb{R}^{2L\times1}$ to denote the joint position parameter vector, where $\boldsymbol{\eta}_l=[z_l,x_l]^{\top}$ is the position of the $l$-th path, consistent with Section~\ref{sec:System Model}. According to \eqref{eq:channel_model}, we define the effective steering vector of the $l$-th path as
\begin{equation}
    \tilde{\mathbf a}_l \triangleq \mathbf a_l(\boldsymbol{\eta}_l)\odot \mathbf d_l .
\end{equation}
Then, the corresponding noise-free compressed observation is
\begin{equation}
    \mathbf s(\boldsymbol{\eta},\mathbf g)
    =
    \sum_{l=1}^{L}\sqrt{P_r}\mathbf W g_l \tilde{\mathbf a}_l,
\end{equation}
where $\mathbf g=[g_1,\dots,g_L]^{\top}$ collects the complex path gains and $\mathbf\Sigma_{\tilde n}=\mathbb E[\tilde{\mathbf n}\tilde{\mathbf n}^H]$ is the effective post-combining noise covariance.
Treating $\mathbf g$ as nuisance parameters, we define
\begin{equation}
    \mathbf B
    =
    \sqrt{P_r}\mathbf W[\tilde{\mathbf a}_1,\dots,\tilde{\mathbf a}_L]
    \in\mathbb C^{N_{\mathrm{RF}}\times L},
\end{equation}
and
\begin{equation}
    \mathbf D
    =
    [\mathbf D_1,\dots,\mathbf D_L]
    \in\mathbb C^{N_{\mathrm{RF}}\times 2L},
    \mathbf D_l
    =
    \sqrt{P_r}g_l\mathbf W
    \left[
    \frac{\partial \tilde{\mathbf a}_l}{\partial z_l},
    \frac{\partial \tilde{\mathbf a}_l}{\partial x_l}
    \right].
\end{equation}
The covariance-weighted orthogonal projector onto the null space of $\mathbf B$ is
\begin{equation}
    \mathbf\Pi_{\mathbf B}^{\perp}
    =
    \mathbf\Sigma_{\tilde n}^{-1}
    -
    \mathbf\Sigma_{\tilde n}^{-1}\mathbf B
    \left(\mathbf B^H\mathbf\Sigma_{\tilde n}^{-1}\mathbf B\right)^{-1}
    \mathbf B^H\mathbf\Sigma_{\tilde n}^{-1},
\end{equation}
which yields the concentrated Fisher information matrix (FIM)
\begin{equation}
    \mathbf F(\boldsymbol{\eta})
    =
    2\Re\!\left\{\mathbf D^H\mathbf\Pi_{\mathbf B}^{\perp}\mathbf D\right\}.
\end{equation}
Therefore, the covariance of any unbiased estimator satisfies
\begin{equation}
    \mathbb E\!\left[
    (\hat{\boldsymbol{\eta}}-\boldsymbol{\eta})
    (\hat{\boldsymbol{\eta}}-\boldsymbol{\eta})^{\top}
    \right]
    \succeq
    \mathbf F^{-1}(\boldsymbol{\eta}).
\end{equation}
Accordingly, the CRLB on the $l$-th path position RMSE is
\begin{equation}
    \mathrm{RMSE}_{l}^{\mathrm{CRLB}}
    \triangleq
    \sqrt{
    \mathrm{tr}\!\left(
    [\mathbf F^{-1}(\boldsymbol{\eta})]_{2l-1:2l,\;2l-1:2l}
    \right)}.
    \label{eq:crlb_rmse_position}
\end{equation}
Finally, the multi-path position bound can be expressed as
\begin{equation}
    \mathrm{RMSE}_{\mathrm{avg}}^{\mathrm{CRLB}}
    \triangleq
    \sqrt{\frac{1}{L}\sum_{l=1}^{L}\left(\mathrm{RMSE}_{l}^{\mathrm{CRLB}}\right)^2 }.
\end{equation}

\subsection{CRLB for Channel Reconstruction}

We define the augmented real-valued parameter vector as
\begin{equation}
    \boldsymbol{\xi}
    =
    [\boldsymbol{\xi}_1^{\top},\dots,\boldsymbol{\xi}_L^{\top}]^{\top}
    \in\mathbb R^{4L\times1},
    \qquad
    \boldsymbol{\xi}_l
    =
    [z_l,x_l,\Re\{g_l\},\Im\{g_l\}]^{\top}.
\end{equation}
The joint FIM associated with $\boldsymbol{\xi}$ can be written as
\begin{equation}
    \mathbf J(\boldsymbol{\xi})
    =
    2\Re\!\left\{
    \left(
    \frac{\partial \mathbf s(\boldsymbol{\xi})}{\partial \boldsymbol{\xi}}
    \right)^H
    \mathbf\Sigma_{\tilde n}^{-1}
    \left(
    \frac{\partial \mathbf s(\boldsymbol{\xi})}{\partial \boldsymbol{\xi}}
    \right)
    \right\}.
\end{equation}
We use $\mathbf G\triangleq \partial\mathbf h/\partial\boldsymbol{\xi}\in\mathbb C^{N\times4L}$ to denote the channel Jacobian. By first-order covariance propagation, the channel estimation error covariance satisfies
\begin{equation}
    \mathbf C_{\mathbf h}
    \succeq
    \mathbf G\mathbf J^{-1}(\boldsymbol{\xi})\mathbf G^H.
\end{equation}
Accordingly, the CRLB for channel reconstruction is
\begin{equation}
    \operatorname{CRLB}_{\mathrm{NMSE}}
    \triangleq
    \frac{\mathrm{tr}(\mathbf C_{\mathbf h})}{\|\mathbf h\|_2^2}.
    \label{eq:crlb_nmse_total}
\end{equation}
}

 \small\bibliographystyle{./bibliography/IEEEtran}
 \bibliography{ref}

@article{nnomp1,
  title   = {{Near-field localization and channel reconstruction for ELAA systems}},
  author  = {Z. Lu and others},
  journal = {IEEE Trans. Wireless Commun.},
  year    = {2024},
  month   = {Jul.},
  number  = {7},
  pages   = {6938--6953},
  volume  = {23}
}

@article{tian_vr,
  title   = {{Low-overhead localization and VR identification for subarray-based ELAA systems}},
  author  = {J. Tian and others},
  journal = {IEEE Wireless Commun. Lett.},
  year    = {2023},
  month   = {May},
  number  = {5},
  pages   = {784--788},
  volume  = {12}
}

@inproceedings{wingloss,
  title     = {{Wing loss for robust facial landmark localisation with convolutional neural networks}},
  author    = {Z. Feng and others},
  booktitle = {Proc. IEEE/CVF Conf. Comput. Vis. Pattern Recognit. (CVPR)},
  year      = {2018},
  month     = {Jun.},
  pages     = {2235--2245}
}

@article{nomp,
  title   = {{Newtonized orthogonal matching pursuit: Frequency estimation over the continuum}},
  author  = {B. Mamandipoor and D. Ramasamy and U. Madhow},
  journal = {IEEE Trans. Signal Process.},
  year    = {2016},
  month   = {Oct.},
  number  = {19},
  pages   = {5066--5081},
  volume  = {64}
}

@article{extendedNOMPbasedTracking,
  title   = {{Tracking FDD massive MIMO downlink channels by exploiting delay and angular reciprocity}},
  author  = {Y. Han and others},
  journal = {IEEE J. Sel. Topics Signal Process.},
  year    = {2019},
  month   = {Sep.},
  number  = {5},
  pages   = {1062--1076},
  volume  = {13}
}

@article{SIGW,
  author  = {M. Cui and L. Dai},
  title   = {{Channel estimation for extremely large-scale MIMO: Far-field or near-field?}},
  journal = {IEEE Trans. Commun.},
  year    = {2022},
  volume  = {70},
  number  = {4},
  pages   = {2663--2677},
  month   = {Dec.}
}

@inproceedings{centernet,
  title     = {{CenterNet: Keypoint triplets for object detection}},
  author    = {K. Duan and others},
  booktitle = {Proc. IEEE/CVF Int. Conf. Comput. Vis. (ICCV)},
  year      = {2019},
  month     = {Nov.},
  pages     = {6568--6577}
}

@article{deepunfolding_hybridBF,
  title   = {{Deep unfolding hybrid beamforming designs for THz massive MIMO systems}},
  author  = {N. T. Nguyen and others},
  journal = {IEEE Trans. Signal Process.},
  year    = {2023},
  pages   = {3788--3803},
  month = {Oct.},
  volume  = {71}
}

@article{hybrid_ris1,
  title   = {{Channel estimation for near-field XL-RIS-aided mmWave hybrid beamforming architectures}},
  author  = {S. Yang and others},
  journal = {IEEE Trans. Veh. Technol.},
  year    = {2023},
  month   = {Aug.},
  number  = {8},
  pages   = {11029--11034},
  volume  = {72}
}

@article{Hungarian,
  author  = {H. W. Kuhn},
  title   = {{The Hungarian method for the assignment problem}},
  journal = {Nav. Res. Logist. Q.},
  year    = {1955},
  volume  = {52},
  pages   = {83--97}
}

@article{keypointCE,
  author  = {M. Li and others},
  title   = {{Keypoint detection empowered near-field user localization and channel reconstruction}},
  journal = {IEEE Trans. Wireless Commun.},
  year    = {2025},
  month   = {Jul.},
  number  = {7},
  pages   = {5519--5533},
  volume  = {24}
}

@article{XL_MIMO_Tutorial,
  title   = {{A tutorial on extremely large-scale MIMO for 6G: Fundamentals, signal processing, and applications}},
  author  = {Zhiqing Wang{\relax} and others},
  journal = {IEEE Wireless Commun.},
  volume  = {31},
  number  = {4},
  year    = {2024},
  month   = {Aug.},
  pages   = {12--20}
}

@article{NF_Localization_Joint_CE,
  title   = {{Near-field user localization and channel estimation for XL-MIMO systems: Fundamentals, recent advances, and outlooks}},
  author  = {H. Lei and others},
  journal = {IEEE Wireless Commun.},
  volume  = {32},
  number  = {4},
  year    = {2025},
  month   = {Aug.},
  pages   = {190--198}
}

@article{CS_Tips_Tricks,
  title   = {{Compressed sensing for wireless communications: Useful tips and tricks}},
  author  = {J. Choi and others},
  journal = {IEEE Commun. Surveys Tuts.},
  volume  = {19},
  number  = {3},
  year    = {2017},
  pages   = {1527--1549}
}

@article{ScattererLocalization_Spherical,
  title   = {{A spherical-wavefront-based scatterer localization algorithm using large-scale antenna arrays}},
  author  = {J. Chen and others},
  journal = {IEEE Commun. Lett.},
  volume  = {20},
  number  = {9},
  year    = {2016},
  month   = {Sep.},
  pages   = {1796--1799}
}

@inproceedings{ScattererLocalization_SAGE,
  title     = {{A scatterer localization method using large-scale antenna array systems based on the SAGE algorithm}},
  author    = {X. Guo and others},
  booktitle = {Proc. IEEE Int. Conf. Antenna Meas. Appl. (CAMA)},
  year      = {2022},
  month     = {Nov.},
  address   = {Cape Town, South Africa},
  pages     = {1--4}
}

@article{Joint_VR_CE_XLMIMO,
  title   = {{Joint visibility region detection and channel estimation for XL-MIMO systems via alternating MAP}},
  author  = {W. Xu and others},
  journal = {IEEE Trans. Signal Process.},
  volume  = {72},
  year    = {2024},
  month   = {Oct.},
  pages   = {4827--4842}
}

@article{Integrated_LocSense_Comm,
  title   = {{An overview on integrated localization and communication towards 6G}},
  author  = {Z. Xiao and Y. Zeng},
  journal = {Sci. China Inf. Sci.},
  volume  = {65},
  number  = {3},
  year    = {2022},
  month   = {Mar.},
  pages   = {131301}
}

@article{Lu_COMST_XLMIMO_Tutorial,
  title   = {{A tutorial on near-field XL-MIMO communications toward 6G}},
  author  = {H. Lu and others},
  journal = {IEEE Commun. Surveys Tuts.},
  volume  = {26},
  number  = {4},
  year    = {2024},
  month   = {4th Quarter},
  pages   = {2213--2257},
  doi     = {10.1109/COMST.2024.3387749}
}

@article{Wang_SCIS_6G_Vision,
  title   = {{Vision, application scenarios, and key technology trends for 6G mobile communications}},
  author  = {Zhe Wang and others},
  journal = {Sci. China Inf. Sci.},
  volume  = {65},
  number  = {5},
  pages   = {1--27},
  year    = {2022},
  doi     = {10.1007/s11432-021-3351-5}
}

@article{Zhu_Subarray_WCL2023,
  title   = {{Sub-array based millimeter wave massive MIMO channel estimation}},
  author  = {X. Zhu and Y. Liu and C.-X. Wang},
  journal = {IEEE Wireless Commun. Lett.},
  volume  = {12},
  number  = {9},
  year    = {2023},
  month   = {Sep.},
  pages   = {1608--1612}
}

@article{Wang_XLMIMO_Mag_2024,
  title   = {{Extremely large-scale MIMO: Fundamentals, challenges, solutions, and future directions}},
  author  = {Z. Wang and others},
  journal = {IEEE Wireless Commun.},
  volume  = {31},
  number  = {3},
  year    = {2024},
  month   = {Jun.},
  pages   = {117--124}
}

@article{Xi_Gridless_WCL2024,
  title   = {{Gridless hybrid-field channel estimation for extra-large aperture array massive MIMO systems}},
  author  = {Y. Xi and others},
  journal = {IEEE Wireless Commun. Lett.},
  volume  = {13},
  number  = {2},
  year    = {2024},
  month   = {Feb.},
  pages   = {496--500},
  doi     = {10.1109/LWC.2023.3333531}
}

@techreport{3GPP3811,
  author      = {{3GPP}},
  title       = {{NR}; Physical channels and modulation},
  institution = {3rd Generation Partnership Project (3GPP)},
  type        = {TS},
  number      = {38.211},
  note        = {Release 18}
}

@techreport{3GPP3813,
  author      = {{3GPP}},
  title       = {{NR}; Physical layer procedures for control},
  institution = {3rd Generation Partnership Project (3GPP)},
  type        = {TS},
  number      = {38.213},
  note        = {Release 18}
}

@article{Zhang_TWC_NFTimeVarying_2026,
  title   = {{Near-Field Time-Varying Channel: Analysis and Tracking}},
  author  = {X. Zhang and H. Zhang and Y. C. Eldar},
  journal = {IEEE Trans. Wireless Commun.},
  volume  = {25},
  year    = {2026},
  pages   = {9280--9294},
  doi     = {10.1109/TWC.2025.3649506}
}

@article{Xu_DynamicSparsity_2024,
  title   = {{Exploiting Dynamic Sparsity for Near-Field Spatial Non-Stationary XL-MIMO Channel Tracking}},
  author  = {W. Xu and others},
  journal = {arXiv preprint arXiv:2412.19475},
  year    = {2024},
  month   = {Dec.}
}

@inproceedings{Yuan_WCNC_XLRIS_2024,
  title     = {{Near-Field Tracking with Extremely Large-Scale RIS: A Sparse Learning Approach}},
  author    = {Y. Yuan and others},
  booktitle = {Proc. IEEE Wireless Commun. Netw. Conf. (WCNC)},
  year      = {2024},
  month     = {Apr.},
  pages     = {1--6},
  doi       = {10.1109/WCNC57260.2024.10571309}
}

@article{Tuo_XLIRS_SparseBayesian_2025,
  title   = {{Near-Field Sparse Bayesian Channel Estimation and Tracking for XL-IRS-Aided Wideband mmWave Systems}},
  author  = {X. Tuo and others},
  journal = {arXiv preprint arXiv:2511.18752},
  year    = {2025},
  month   = {Nov.}
}

@article{Zhao_TCOM_AngleTracking_2017,
  author  = {J. Zhao and others},
  journal = {IEEE Trans. Wireless Commun.}, 
  title   = {Angle Domain Hybrid Precoding and Channel Tracking for Millimeter Wave Massive {MIMO} Systems}, 
  year    = {2017},
  volume  = {16},
  number  = {10},
  pages   = {6868--6880},
  doi     = {10.1109/TWC.2017.2732405}
}

@inproceedings{Zhao_TWC_STBEM_2018,
  author    = {J. Zhao and others},
  booktitle = {Proc. IEEE Int. Conf. Commun. (ICC)}, 
  title     = {Channel Tracking for Massive {MIMO} Systems With Spatial-Temporal Basis Expansion Model}, 
  year      = {2017},
  pages     = {1--5},
  doi       = {10.1109/ICC.2017.7996914}
}

@article{Guerra_TSP_NFTracking_2021,
  author  = {A. Guerra and others},
  journal = {IEEE Trans. Signal Process.}, 
  title   = {Near-Field Tracking With Large Antenna Arrays: Fundamental Limits and Practical Algorithms}, 
  year    = {2021},
  volume  = {69},
  pages   = {5723--5738},
  month   = {Oct.},
  doi     = {10.1109/TSP.2021.3115100}
}

@article{Chen_Arxiv_HybridTracking_2025,
  title   = {{Near-Field Position and Orientation Tracking With Hybrid ELAA Architecture}},
  author  = {L. Chen and X. Yuan and Y.-J. A. Zhang},
  journal = {arXiv preprint arXiv:2512.17274},
  year    = {2025},
  month   = {Dec.}
}

@article{Lee_TCOM_ChannelEstimation_2016,
  author  = {J. Lee and G.-T. Gil and Y. H. Lee},
  journal = {IEEE Trans. Commun.}, 
  title   = {Channel estimation via orthogonal matching pursuit for hybrid {MIMO} systems in millimeter wave communications}, 
  year    = {2016},
  volume  = {64},
  number  = {6},
  pages   = {2370--2386},
  doi     = {10.1109/TCOMM.2016.2557791}
}

@book{Kay_Estimation_1993,
  author    = {Steven M. Kay},
  title     = {Fundamentals of Statistical Signal Processing, Volume I: Estimation Theory},
  publisher = {Prentice-Hall},
  address   = {Englewood Cliffs, NJ, USA},
  year      = {1993}
}

@inproceedings{cornernet,
  title     = {{CornerNet: Detecting Objects as Paired Keypoints}},
  author    = {H. Law and J. Deng},
  booktitle = {Proc. Eur. Conf. Comput. Vis. (ECCV)},
  year      = {2018},
  month     = {Sep.},
  pages     = {734--750}
}

@article{objectsaspoints,
  title   = {{Objects as Points}},
  author  = {X. Zhou and others},
  journal = {arXiv preprint arXiv:1904.07850},
  year    = {2019},
  month   = {Apr.}
}

@inproceedings{hrnetpose,
  title     = {{Deep High-Resolution Representation Learning for Human Pose Estimation}},
  author    = {K. Sun and others},
  booktitle = {Proc. IEEE/CVF Conf. Comput. Vis. Pattern Recognit. (CVPR)},
  year      = {2019},
  month     = {June},
  pages     = {5693--5703}
}

@inproceedings{centertrack,
  title     = {{Tracking Objects as Points}},
  author    = {X. Zhou and others},
  booktitle = {Proc. Eur. Conf. Comput. Vis. (ECCV)},
  year      = {2020},
  month     = {Aug.},
  pages     = {474--490}
}

@inproceedings{centerpoint,
  title     = {{Center-Based 3D Object Detection and Tracking}},
  author    = {T. Yin and others},
  booktitle = {Proc. IEEE/CVF Conf. Comput. Vis. Pattern Recognit. (CVPR)},
  year      = {2021},
  month     = {June},
  pages     = {11784--11793}
}

@inproceedings{nms,
  title     = {{Efficient Non-Maximum Suppression}},
  author    = {A. Neubeck and L. Van Gool},
  booktitle = {Proc. Int. Conf. Pattern Recognit. (ICPR)},
  year      = {2006},
  month     = {Aug.},
  pages     = {850--855}
}

\vspace{12pt}
\color{red}

\end{document}